\documentclass[11pt,english]{article}
\usepackage[latin9]{inputenc}
\usepackage{geometry}
\usepackage{babel}
\usepackage{float}
\usepackage{amsmath}
\usepackage{amssymb}
\usepackage{graphicx}
\usepackage{setspace}
\usepackage[authoryear]{natbib}
\usepackage[unicode=true,pdfusetitle,
 bookmarks=true,bookmarksnumbered=false,bookmarksopen=false,
 breaklinks=false,pdfborder={0 0 0},pdfborderstyle={},backref=false,colorlinks=false]
 {hyperref}

\makeatletter

\newcommand{\lyxdot}{.}

\hypersetup{colorlinks=true,linkcolor=blue,citecolor=blue,urlcolor=blue}
\usepackage{diagbox}
\usepackage{algorithm}
\usepackage{algorithmic}

\usepackage{amsmath}

\@ifundefined{showcaptionsetup}{}{%
 \PassOptionsToPackage{caption=false}{subfig}}
\usepackage{subfig}
\makeatother

\begin{document}
\title{A probabilistic model for missing traffic volume reconstruction based
on data fusion}
\author{Xintao Yan\textsuperscript{a}, Yan Zhao\textsuperscript{b}, Henry
X. Liu\textsuperscript{a,c,}\thanks{Corresponding author.}\date{}}

\maketitle
\begin{singlespace}
\textsuperscript{a}\textit{Department of Civil and Environmental
Engineering, University of Michigan, Ann Arbor, MI, USA}

\textsuperscript{b}\textit{Department of Mechanical Engineering,
University of Michigan, Ann Arbor, MI, USA}

\textsuperscript{c}\textit{University of Michigan Transportation
Research Institute, University of Michigan, Ann Arbor, MI, USA}
\end{singlespace}

\section*{Abstract}

Traffic volume information is critical for intelligent transportation
systems. It serves as a key input to transportation planning, roadway
design, and traffic signal control. However, the traffic volume data
collected by fixed-location sensors, such as loop detectors, often
suffer from the missing data problem and low coverage problem. The
missing data problem could be caused by hardware malfunction. The
low coverage problem is due to the limited coverage of fixed-location
sensors in the transportation network, which restrains our understanding
of the traffic at the network level. To tackle these problems, we
propose a probabilistic model for traffic volume reconstruction by
fusing fixed-location sensor data and probe vehicle data. We apply
the probabilistic principal component analysis (PPCA) to capture the
correlations in traffic volume data. An innovative contribution of
this work is that we also integrate probe vehicle data into the framework,
which allows the model to solve both of the above-mentioned two problems.
Using a real-world traffic volume dataset, we show that the proposed
method outperforms state-of-the-art methods for the extensively studied
missing data problem. Moreover, for the low coverage problem, which
cannot be handled by most existing methods, the proposed model can
also achieve high accuracy. The experiments also show that even when
the missing ratio reaches 80\%, the proposed method can still give
an accurate estimate of the unknown traffic volumes with only 10\%
probe vehicle penetration rate. The results validate the effectiveness
and robustness of the proposed model and demonstrate its potential
for practical applications. \textbf{\textit{}}\\
\textbf{\textit{}}\\
\textbf{\textit{Keywords}}\textit{: }Traffic volume, Probe vehicle,
Data fusion, Probabilistic principle component analysis

\noindent \pagebreak{}

\section{Introduction \label{sec:Introduction}}

Traffic volume information plays a critical role in transportation
planning, roadway design, and traffic signal control. In conventional
transportation systems, traffic volumes are primarily measured by
fixed-location sensors, such as loop detectors \citep{guo2019urban}.
Although widely applied, loop detectors have the following two significant
drawbacks. The first drawback is that the collected data often contain
missing values, which might be caused by hardware malfunction. Another
drawback of loop detectors is that they usually only cover a small
subset of links in a transportation network, due to the high installation
and maintenance costs \citep{yoon2007-maintenance-cost,zhan2016citywide}.
Therefore, loop detectors usually can only measure very limited traffic
volume information, which restrains our understanding of the traffic
at the network level.

To tackle the first problem, i.e., the missing data problem, abundant
literature applied data imputation methods to loop detector data.
The key idea of these methods is to exploit the spatiotemporal correlation
of the traffic volume data. The methods can be roughly divided into
three categories. The first category is based on principal component
analysis (PCA), which includes the applications of probabilistic PCA
(PPCA), kernel probabilistic PCA (KPPCA), Bayesian PCA (BPCA), and
their variants \citep{qu2008bpca,qu2009ppca,ilin2010practical,li2013efficient}.
These probabilistic methods try to find and exploit the low-rank structure
of the traffic volume data for missing value imputation. The second
category is based on the matrix (tensor) completion. The methods in
this category usually represent traffic volume data as a matrix (tensor)
and impute the traffic data by matrix (tensor) decomposition techniques
\citep{tan2013tensor,asif2016matrix,ran2016tensor,goulart2017traffic,chen2019missing,chen2019bayesian}.
The third category mainly focuses on data-driven machine learning
methods, including neural networks \citep{duan2016efficient,zhuang2018innovative,chen2019traffic,li2020real-Ban-2},
k-nearest neighbors \citep{tak2016data}, and CoKriging methods \citep{bae2018missing}.

When it comes to the second drawback of loop detectors, the low coverage
problem, using solely loop detector data is usually not sufficient
to solve the problem. If loop detectors are not installed at the location
where the traffic volume information is of our interest, the data
imputation methods introduced above could not be applied, because
all of the methods require at least one observed data point for each
location. Recently, a wide range of methods from the perspective of
probability theory and statistics have been proposed for estimating
traffic volumes using probe vehicle data. \citet{zheng2017estimating}
modeled vehicle arrivals at an intersection following a time-varying
Poisson process and estimate the traffic volumes using maximum likelihood
estimation (MLE). \citet{wang2019-Ind} constructed a Bayesian network
to capture the relationship between vehicle arrival processes and
the timing information in probe vehicle trajectory data. The traffic
volume can be calculated based on the inferred traffic parameters
from the Bayesian network by applying the expectation-maximization
(EM) algorithm. Unlike previous work that only considered isolated
intersections, \citet{luo2019arterial} improved the estimation accuracy
by considering the information from adjacent intersections. The traffic
volume can also be estimated by scaling up the probe vehicle volume
with the penetration rate. A recent study by \citet{wong2019estimation}
proposed a novel method that provides an unbiased estimator for probe
vehicle penetration rate. \citet{zhao2019volume,zhao2019various}
also developed a series of estimators for probe vehicle penetration
rate, which can be further used for traffic volume and queue length
estimation. However, these direct scaling methods cannot handle the
cases when there are no observations of probe vehicles, which could
be caused by low market penetration or fine estimation time granularity.
In general, under the low penetration rate, using probe vehicle data
alone results in a trade-off between time granularity and estimation
accuracy.

Combining loop detector data and probe vehicle data could potentially
solve the two problems at the same time. On the one hand, despite
the low coverage, when loop detectors function well, they could give
the complete vehicle counts at discrete locations. On the other hand,
although the penetration rate of probe vehicles is low currently,
probe vehicles usually have much broader coverage and do not have
the maintenance issues. Therefore, the fusion of the two data sources
makes their respective advantages complementary to each other. A few
recent studies attempted to fuse the two data sources. \citet{cui2017mining}
estimated the unknown traffic volumes by applying compressive sensing
techniques. The correlation of traffic volumes in adjacent time slots
was captured by a Toeplitz matrix; the correlation of traffic volumes
in nearby locations was learned by fitting linear regression models
to probe vehicle counts. Besides the two data sources, \citet{zhan2016citywide}
also include points of interest (POI) data and meteorology data to
develop a hybrid framework that extracted some high-level features
from calibrated fundamental diagrams and estimated traffic volumes
by machine learning techniques. \citet{meng2017city} modeled the
spatiotemporal correlation of traffic volumes by a multi-layer affinity
graph.

In this paper, we first propose a general probabilistic framework
for traffic state estimation problems. Based on the framework, we
propose a data fusion method to simultaneously address the two challenges
of traffic volume reconstruction, namely, the missing data problem
and the low coverage problem. In doing so, we adopt PPCA to capture
the low-rank structure of traffic volume data. Besides the spatiotemporal
correlations contained in loop detector data, the proposed model also
captures the sampling process of probe vehicle data and thus allows
us to impute the missing traffic volumes more accurately and robustly.
Most importantly, for locations that are not covered by loop detectors,
the method can still accurately reconstruct the traffic volumes, whereas
this scenario is very challenging for existing methods.

The main contributions of this paper are fourfold: (1) we propose
a general probabilistic framework for traffic state estimation problems;
(2) we propose a data fusion method that exploits both fixed-location
sensor data and probe vehicle data to capture the spatiotemporal correlations
in traffic volumes; (3) the proposed method can estimate the traffic
volume for some locations where no loop detectors are installed, as
long as there are probe vehicle data; (4) in terms of the extensively
studied missing traffic volume problem, the proposed method outperforms
the existing methods, as the results on the real-world dataset suggest.

The rest of this paper is organized as follows. In Section \ref{sec:General probabilistic framework},
we introduce the general probabilistic framework for traffic state
estimation problems. A stream of literature that conforms to this
framework is discussed in detail to demonstrate the universality of
the proposed framework. In Section \ref{sec:Probabilistic-models},
the probabilistic models for traffic volume data in probe vehicle
environments are given. Section \ref{sec:Data-fusion-by-PPCA} presents
how to model the correlations in traffic volumes and how to incorporate
loop detector data and probe vehicle data into the model. The model
parameters can be estimated by solving a maximum likelihood estimation
problem using the EM algorithm. In Section \ref{sec:Case-studies},
we evaluate our method by extensive experiments under different settings
using a public traffic flow dataset collected from Portland, Oregon.
We also compare the performance of the proposed method with existing
methods and demonstrate that our method outperforms them, benefiting
from the fusion of different data sources. Finally, Section \ref{sec:Conclusions}
concludes the study and suggests some future directions.

\section{A general probabilistic framework for traffic state estimation problems
\label{sec:General probabilistic framework}}

Traffic states represent traffic conditions at a given location and
time. Traffic state variables include traffic volume, traffic speed, traffic density, travel time, and queue length, etc. Before looking into traffic volume estimation, we first
provide a probabilistic view of the general traffic state estimation
problem. The proposed method for the missing traffic volume reconstruction
will follow this general probabilistic framework.

\begin{figure}[H]
\begin{centering}
\includegraphics[width=0.8\textwidth]{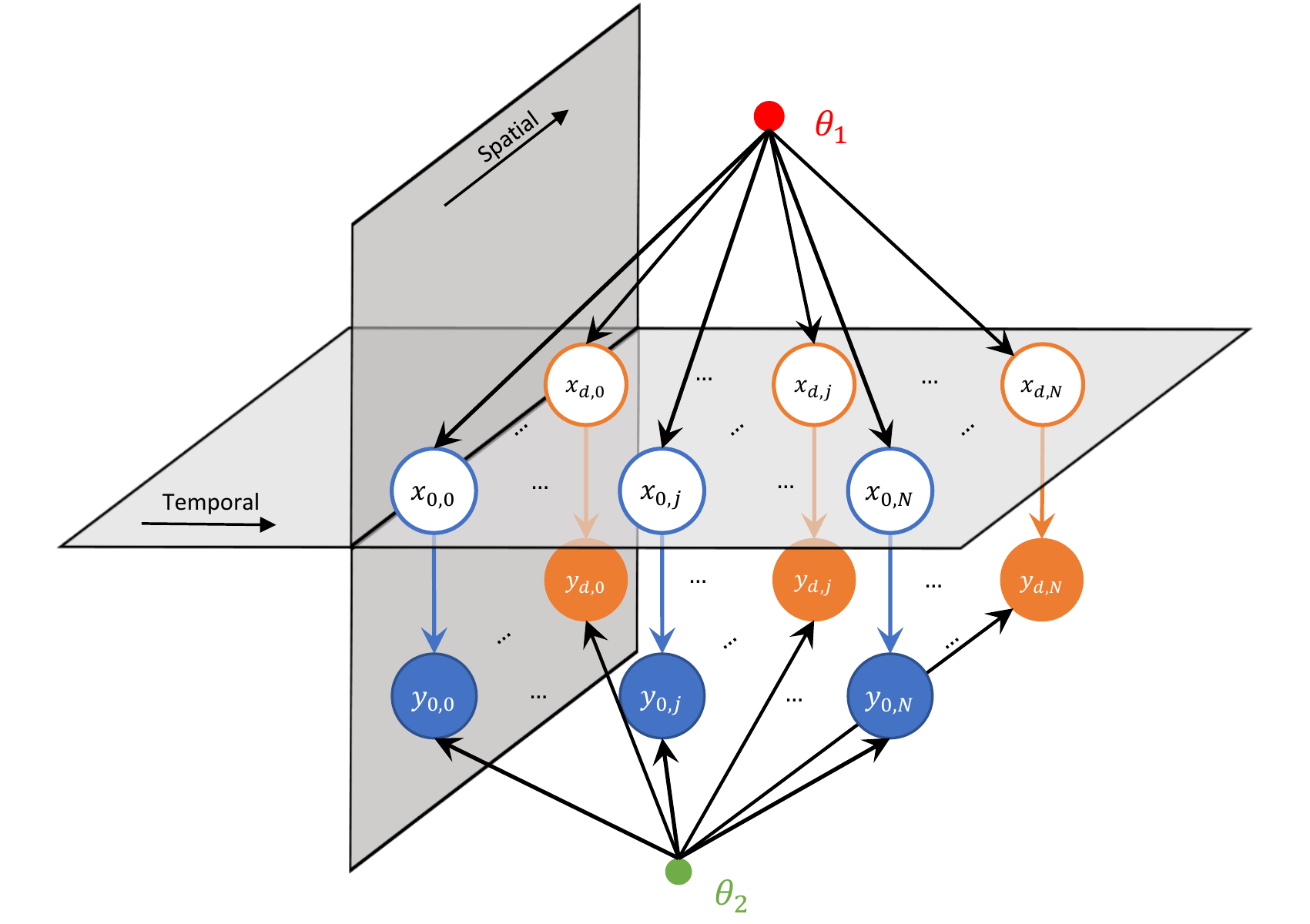}
\par\end{centering}
\caption{The general Bayesian network for traffic state estimation problem.
\label{fig: general-framework}}
\end{figure}

Within a transportation network, traffic states in different locations
and times are inherently spatially and temporally correlated. To estimate
traffic states, we use observations obtained from either fixed-location
sensors, for example, loop-detectors and cameras, or moving sensors
such as the probe vehicles. Here we propose a general Bayesian network
for traffic state estimation problems as shown in Figure \ref{fig: general-framework}.
The Bayesian network can model spatial and temporal dependencies between
traffic states and also capture the conditional dependencies between
traffic states and observations.

The traffic state of location $i$ at time $j$ is denoted by $x_{ij}$
as shown by the circle on the upper layer in the figure. $d$ and $N$ denote the number of locations and time slots being considered. The corresponding
observation from the probe vehicles of each traffic state $x_{ij}$
is denoted by $y_{ij}$, which is represented by the circle on the
lower layer. The traffic states depend on a set of parameters $\theta_{1}$,
and the probe vehicle observations depend on both the traffic states
and another set of parameters $\theta_{2}$. The parameters are shown
as the small solid circles in the figure. The probabilistic framework
represents a general traffic state estimation methodology. In order
to demonstrate the universality of the framework, in the following
paragraphs, we will discuss in detail some literature that conforms
to this framework.

The first category of literature imposes independent assumptions,
which implies the traffic states of different locations at different
times are independent with each other. As an example, \citet{comert2009queue}
estimated the queue length in each cycle at an isolated and undersaturated
intersection using probe vehicle data. The traffic state variable is the queue
length, and the observation from probe vehicle data is the position
of the last probe vehicle in the queue. Vehicle arrivals are assumed
to follow a Poisson process, therefore, the parameter $\theta_{1}$
is the Poisson arrival rate. Given the queue length of a cycle, the
observation also depends on the probe vehicle penetration rate which
serves as parameter $\theta_{2}$. Similarly, studies by \citet{zheng2017estimating},
\citet{zhao2019various}, and \citet{wang2019-Ind} also fall into
this category.

The second category of literature includes those considering the temporal
correlation (or spatial correlation) alone. \citet{aljamal2020real-KF2-Journal}
proposed a Kalman filter (KF) based method to estimate the traffic
state using probe vehicle data. The traffic state is the number of
vehicles traversing the approach, and the corresponding probe vehicle
observation is the probe vehicle travel time on the approach. The
system dynamics are modeled as a linear system where the temporal
correlation of traffic states is considered and incorporated. Therefore,
parameter $\theta_{1}$ includes transition matrices of the model.
For parameter $\theta_{2},$ it includes the measurement matrices,
covariance of the measurement Gaussian noise, and the penetration
rate. Besides the KF-based method, \citet{wang2015-HMM} proposed
a hidden Markov model (HMM) based method to estimate road segment's
travel speed using probe vehicle data, and it also belongs to this
category.

The last category of literature involves those considering both spatial
and temporal correlations. In addition to the queue length and traffic
volume discussed earlier, the link travel time is studied in \citet{herring2010-ST2}.
The traffic state of each link is represented by a discrete congestion
level. For a given congestion level, the travel time distribution
of that link is assumed to follow a Gaussian distribution. To capture
the spatiotemporal correlation, the authors assume the state of a
link at the next timestep depends on the states of the spatial neighbors
in the current timestep. The observations are the link travel time
obtained from the probe vehicle data. Therefore, parameter $\theta_{1}$
involves the state transition probability matrix and initial state
probability. Parameter $\theta_{2}$ includes the mean and variance
of the link travel time Gaussian distribution given different congestion
levels. Similarly, the spatiotemporal correlation is also considered
in \citet{chen2019-KF} to estimate the occupancy of each road segment,
and this method also falls into this category.

In summary, the proposed framework illustrated in Figure \ref{fig: general-framework}
represents a generic methodology for traffic state estimation problems.
Based on the methodology, in the next sections, we will propose a
probabilistic model for traffic volume reconstruction by fusing fixed-location
sensor data and probe vehicle data.

\section{Probabilistic models for traffic volume data in probe vehicle environments
\label{sec:Probabilistic-models}}

Before delving into traffic volume reconstruction, we first describe
the probabilistic models of whole-population traffic volumes and probe
vehicle traffic volumes. By whole-population traffic volume, we refer
to the total number of vehicles passing through a certain location
at a certain time interval to distinguish it from the probe vehicle
traffic volume. These probabilistic models are the traffic state and
probe vehicle observation components in the framework that we introduced
in the previous section. They serve as the foundation of the proposed
data fusion method.

\subsection{Distribution of whole-population traffic volumes based on the PPCA
model\label{subsec:Basic-PPCA-framework}}

Following the same notation as in Section \ref{sec:General probabilistic framework},
for a specific time-of-day (TOD), we represent the traffic volumes
of $d$ locations in $N$ days by a matrix $X\in\mathbb{R}^{d\times N}$,
of which the element $x_{ij}$ represents the traffic volume at location
$i$ on day $j$. Many studies have shown that the traffic volume
data have strong spatiotemporal correlations and contain low-rank
structures \citep{qu2009ppca,tan2013tensor,coogan2017traffic,feng2018better}.
We apply the PPCA model proposed by \citet{tipping1999probabilistic}
and \citet{roweis1999unifying} to capture the low-rank structure
of the traffic volume data.

The PPCA model is a probabilistic model that generalizes PCA. The
PPCA model assumes that the $d$-dimensional sample vector $x_{n}$,
which is the $n$th column of $X$, depends on an $r$-dimensional
latent vector $t_{n}$ through the following linear-Gaussian format
\begin{equation}
x_{n}=\Lambda t_{n}+\mu_{x}+\epsilon_{n},
\end{equation}
where the latent vector $t_{n}$ is assumed to follow the multivariate
Gaussian distribution $\mathcal{N}\left(0,I\right)$, and $\Lambda$
is a $d\times r$ projection matrix. $\mu_{x}$ is a vector that captures
the mean of all samples, and $\epsilon_{n}$ is a $d$-dimensional
isotropic Gaussian noise satisfying $\epsilon_{n}\sim\mathcal{N}\left(0,\sigma^{2}I\right)$,
where $\sigma^{2}$ is the variance of the noise in each dimension.
The intuition behind the formulation is that, with $r\ll d$, the
original $d$-dimensional sample data can be represented in a more
sparse way by mapping an $r$-dimensional latent variable in the latent
variable space to the sample data space using the projection matrix
$\Lambda$. According to the model, the distribution of $x_{n}$ is
\begin{equation}
x_{n}\sim\mathcal{N}\left(\mu_{x},\Lambda\Lambda^{T}+\sigma^{2}I\right).\label{eq:traffic-state-distribution}
\end{equation}
Eq. (\ref{eq:traffic-state-distribution}) suggests that the distribution
of traffic states depends on the parameter $\theta_{1}=\{\Lambda,\mu_{x},\sigma^{2}\}$,
which includes the projection matrix, the mean vector, and the variance
of the isotropic Gaussian noise.

\subsection{Distribution of probe vehicle traffic volumes}

The probe vehicle traffic volume represents the number of probe vehicles
passing by a location in a specific period. Similarly, for a specific
TOD, we represent the probe vehicle traffic volumes of $d$ locations
in $N$ days by a matrix $Y$, which shares the same size as $X$.
For location $i$ and day $j$, the probe vehicle traffic volume $y_{ij}$
is a fraction of the whole-population traffic volume $x_{ij}$.

We assume that the penetration rate of probe vehicles at each of the
studied locations in the studied TOD is the same, denoted by $p$.
We also assume probe vehicles are randomly mixed with regular vehicles.
Based on the assumptions, given the penetration rate $p$ and the
whole-population traffic volume $x_{ij}$, the traffic volume of probe
vehicles at location $i$ on day $j$ follows the binomial distribution
$y_{ij}|x_{ij}\sim\mathcal{B}\left(x_{ij},p\right)$. The binomial
distribution can be well approximated by a Gaussian distribution when
$x_{ij}$ is large \citep{shiryayev1984probability}, which is adequate
in our case. Therefore, the probe vehicle volume vector $y_{n}$ (the
$n$th column of $Y$) approximately follows the Gaussian distribution
\begin{equation}
y_{n}\mid x_{n}\sim\mathcal{N}\left(x_{n}p,{\rm diag}\left(x_{n}p\left(1-p\right)\right)\right).
\end{equation}
Another reason we approximate the binomial distribution using a Gaussian
is that the PPCA framework applies to continuous random variables,
whereas real-world traffic volumes are integer values. The Gaussian
approximation makes it easy to consider the loop detector data and
probe vehicle data together.

We also applied another approximation to the distribution of $y_{n}$,
for mathematical simplification. We substitute the average traffic
volume $\bar{x}=\sum_{i=1}^{N}x_{n}/N$ for the traffic volume $x_{n}$
in the variance of the distribution and decouple the mean and covariance
by replacing $p\left(1-p\right)$ with $\eta^{2}$. Consequently,
the probability distribution of the probe vehicle traffic volume is
expressed as
\begin{equation}
y_{n}\mid x_{n}\sim\mathcal{N}\left(x_{n}p,\bar{x}\eta^{2}\right).\label{eq: Probe_Gau_approx}
\end{equation}
Therefore, the probe vehicle observation depends on the parameter
$\theta_{2}=\{p,\eta^{2}\}$, where $p$ and $\eta^{2}$ denote the
probe vehicle penetration rate and the decoupled variance, respectively.

\section{Traffic volume reconstruction by data fusion \label{sec:Data-fusion-by-PPCA}}

After modeling the distributions of whole-population traffic volumes
and probe vehicle traffic volumes, in this section, we show how to
fuse the loop detector data and probe vehicle data and how to infer
the model parameters by solving an MLE problem. Once the parameters
in the model are obtained, we will be able to reconstruct the unknown
traffic volumes easily.

\subsection{PPCA based data fusion \label{subsec:Traffic-volume-reconstruction}}

This paper focuses on the reconstruction of traffic volumes in two
specific scenarios, i.e., the missing data scenario and the low coverage
scenario. In the missing data scenario, some of the entries in $X$
might be missing due to loop detector malfunction or communication
failure. The missing entries show a random pattern. In the low coverage
scenario, loop detectors are not installed in some locations of our
interest, and the entries of the corresponding rows in $X$ will be
empty. Our goal is to reconstruct the missing values in $X$ by fusing
the non-missing values in $X$ and the probe vehicle data $Y$.

In this study, we incorporate the probe vehicle data into the PPCA
model and propose a PPCA-based data fusion (PPCA-DF) model. Following
the notation in \citet{marlin2008missing}, when there are missing
elements in traffic volume data $x_{n}$, we divide $x_{n}$ into
two parts $x_{n}^{m}$ and $x_{n}^{o}$, where $x_{n}^{m}$ refers
to the missing part and $x_{n}^{o}$ represents the non-missing part.
The available (non-missing) traffic volume data and probe vehicle
data serve as the model input. For conciseness, we denote the collection
of model parameters by $\theta=\left\{ \Lambda,\mu_{x},\sigma^{2},p,\eta^{2}\right\} $.
Given the non-missing traffic volume and probe vehicle volume data,
the log-likelihood function of $\theta$ can be expressed as 
\begin{align}
\log\mathcal{L}\left(\theta;x_{n}^{o},y_{n}\right) & =\sum_{n=1}^{N}\log P_{\theta}\left(x_{n}^{o},y_{n}\right)\label{eq:log-likelihood}\\
 & =\sum_{n=1}^{N}\log\left(\int P_{\theta}\left(x_{n}^{m},x_{n}^{o},y_{n}\right)dx_{n}^{m}\right)\label{eq:log-likelihood2}\\
 & =\sum_{n=1}^{N}\log\left(\int\int P_{\theta}\left(x_{n}^{m},x_{n}^{o},y_{n},t_{n}\right)dx_{n}^{m}dt_{n}\right).\label{eq:log-likelihood3}
\end{align}
The objective function of the PPCA-DF model is to maximize the log-likelihood
function. For the convenience of the solving process, we introduce
the latent vector $t_{n}$ to the objective function as shown in Eq.
(\ref{eq:log-likelihood3}).

The complete-data likelihood function in the marginal log-likelihood
function (\ref{eq:log-likelihood3}) can be expressed as 
\begin{align}
P_{\theta}\left(x_{n}^{m},x_{n}^{o},y_{n},t_{n}\right) & =P_{\theta}\left(x_{n},y_{n},t_{n}\right)\\
 & =P_{\theta}\left(t_{n}\right)P_{\theta}\left(x_{n}\rvert t_{n}\right)P_{\theta}\left(y_{n}\rvert x_{n},t_{n}\right)\\
 & =P_{\theta}\left(t_{n}\right)P_{\theta}\left(x_{n}\rvert t_{n}\right)P_{\theta}\left(y_{n}\rvert x_{n}\right).\label{eq:complete_llh_abs}
\end{align}
 The last step in the derivation is because $y_{n}$ is independent
of $t_{n}$ given $x_{n}$. The probability density functions of $t_{n}$,
$x_{n}\rvert t_{n}$, and $y_{n}\rvert x_{n}$ under parameter $\theta$
are

\begin{align}
P_{\theta}\left(t_{n}\right) & =\left(2\pi\right)^{-\frac{r}{2}}e^{-\frac{1}{2}t_{n}^{T}t_{n}},\\
P_{\theta}\left(x_{n}\rvert t_{n}\right) & =\left(2\pi\sigma^{2}\right)^{-\frac{d}{2}}e^{-\frac{1}{2\sigma^{2}}\left(x_{n}-\Lambda t_{n}-\mu_{x}\right)^{T}\left(x_{n}-\Lambda t_{n}-\mu_{x}\right)},\\
P_{\theta}\left(y_{n}\rvert x_{n}\right) & =\frac{1}{\sqrt{\text{\ensuremath{\Pi_{i=1}^{d}\bar{x}_{i}}}}}\left(2\pi\eta^{2}\right)^{-\frac{d}{2}}e^{-\frac{1}{2}\left(y_{n}-px_{n}\right)^{T}\left[{\rm diag}\left(\bar{x}\eta^{2}\right)\right]^{-1}\left(y_{n}-px_{n}\right)},\label{eq:yn_xn}
\end{align}
respectively. Please note that $\bar{x}$ in Eq. (\ref{eq:yn_xn})
is prior information that can be obtained by averaging the non-missing
values.

By substituting these probability density functions into Eq. (\ref{eq:complete_llh_abs}),
the complete-data log-likelihood function can be expressed as
\begin{align}
\log P_{\theta}\left(x_{n}^{m},x_{n}^{o},y_{n},t_{n}\right) & =-\frac{\left(r+2d\right)}{2}\log\left(2\pi\right)-\frac{1}{2}t_{n}^{T}t_{n}-\frac{d}{2}\log\left(\sigma^{2}\right)\nonumber \\
 & -\frac{1}{2\sigma^{2}}\left(x_{n}-\Lambda t_{n}-\mu_{x}\right)^{T}\left(x_{n}-\Lambda t_{n}-\mu_{x}\right)-\frac{1}{2}\sum_{i=1}^{d}\log\left(\bar{x}_{i}\eta^{2}\right)\nonumber \\
 & -\frac{1}{2}\left(y_{n}-px_{n}\right)^{T}\left[{\rm diag}\left(\bar{x}\eta^{2}\right)\right]^{-1}\left(y_{n}-px_{n}\right).\label{eq:complete_llh}
\end{align}
Substituting Eq. (\ref{eq:complete_llh}) into Eq. (\ref{eq:log-likelihood3})
gives a non-concave objective function of the maximum likelihood estimation
problem. Therefore, we apply the EM algorithm \citep{dempster1977maximum}
to solve it.

\subsection{EM algorithm}

\subsubsection{E-step}

The goal of the EM algorithm is to find maximum likelihood solutions
for models having latent variables \citep{bishop2006pattern}. In
the E-step, we evaluate the expectation of the complete-data log-likelihood
function under the posterior distribution of the latent variables
given the current estimate $\theta^{(k)}$, where the superscript
$k$ is the index denoting for current iteration. Mathematically,
the expectation can be expressed as $\mathbb{E}_{t_{n},x_{n}^{m}\rvert x_{n}^{o},y_{n},\theta^{(k)}}\left[\log\mathcal{L}\left(\theta;x_{n},y_{n},t_{n}\right)\right]$,
where $\log\mathcal{L}\left(\theta;x_{n},y_{n},t_{n}\right)={\rm log}P_{\theta}\left(x_{n},y_{n},t_{n}\right)$.

To get the probability density function of the posterior distribution
under parameter $\theta^{(k)}$, we first derive the joint distribution
of $x_{n},y_{n}$ and $t_{n}$, which is 
\begin{equation}
\left(x_{n},y_{n},t_{n}\right);\theta^{(k)}\sim\mathcal{N}\left(\left[\begin{array}{c}
\mu_{x}^{(k)}\\
\mu_{y}^{(k)}\\
0
\end{array}\right],\left[\begin{array}{ccc}
\Sigma_{x_{n}x_{n}}^{(k)} & \Sigma_{x_{n}y_{n}}^{(k)} & \Sigma_{x_{n}t_{n}}^{(k)}\\
\Sigma_{y_{n}x_{n}}^{(k)} & \Sigma_{y_{n}y_{n}}^{(k)} & \Sigma_{y_{n}t_{n}}^{(k)}\\
\Sigma_{t_{n}x_{n}}^{(k)} & \Sigma_{t_{n}y_{n}}^{(k)} & \Sigma_{t_{n}t_{n}}^{(k)}
\end{array}\right]\right),
\end{equation}
where the covariance matrix can be expressed as{\small{}
\begin{align}
\left[\begin{array}{ccc}
(\sigma^{2})^{\left(k\right)}I+\Lambda^{\left(k\right)}\left(\Lambda^{\left(k\right)}\right)^{T} & p^{\left(k\right)}\left((\sigma^{2})^{\left(k\right)}I+\Lambda^{\left(k\right)}\left(\Lambda^{\left(k\right)}\right)^{T}\right) & \Lambda^{\left(k\right)}\\
p^{\left(k\right)}\left((\sigma^{2})^{\left(k\right)}I+\Lambda^{\left(k\right)}\left(\Lambda^{\left(k\right)}\right)^{T}\right)^{T} & {\rm diag}\left(\bar{x}(\eta^{2})^{\left(k\right)}\right)+\left(p^{\left(k\right)}\right)^{2}\left((\sigma^{2})^{\left(k\right)}I+\Lambda^{\left(k\right)}\left(\Lambda^{\left(k\right)}\right)^{T}\right) & p^{\left(k\right)}\Lambda^{\left(k\right)}\\
\left(\Lambda^{\left(k\right)}\right)^{T} & p^{\left(k\right)}\left(\Lambda^{\left(k\right)}\right)^{T} & I
\end{array}\right].
\end{align}
}Then, according to the Gaussian conditional distribution formula,
the conditional distribution of the latent variables $x_{n}^{m}$
and $t_{n}$ given the observed data $x_{n}^{o}$ and $y_{n}$ is
still Gaussian. For conciseness, we denote the distribution of $\left(t_{n},x_{n}^{m}\rvert x_{n}^{o},y_{n};\theta^{(k)}\right)$
by $q_{n}^{(k)}\left(t_{n},x_{n}^{m}\right)$, that is
\begin{equation}
q_{n}^{(k)}\left(t_{n},x_{n}^{m}\right):t_{n},x_{n}^{m}\rvert x_{n}^{o},y_{n};\theta^{(k)}\sim\mathcal{N}\left(\left[\begin{array}{c}
\mu_{t_{n}\rvert x_{n}^{o},y_{n}}^{(k)}\\
\mu_{x_{n}^{m}\rvert x_{n}^{o},y_{n}}^{(k)}
\end{array}\right],\left[\begin{array}{cc}
\Sigma_{t_{n}\rvert x_{n}^{o},y_{n}}^{(k)} & \Sigma_{t_{n}x_{n}^{m}\rvert x_{n}^{o},y_{n}}^{(k)}\\
\left(\Sigma_{t_{n}x_{n}^{m}\rvert x_{n}^{o},y_{n}}^{(k)}\right)^{T} & \Sigma_{x_{n}^{m}\rvert x_{n}^{o},y_{n}}^{(k)}
\end{array}\right]\right).\label{eq:derive5expectations}
\end{equation}

Finally, we evaluate the expected complete-data log-likelihood function
under the posterior distribution $q_{n}^{(k)}(t_{n},x_{n}^{m})$,
which is 
\begin{align}
\mathbb{E}_{q_{n}^{(k)}}\left[\mathcal{\log L}\left(\theta;x_{n},y_{n},t_{n}\right)\right] & =\int\int q_{n}^{(k)}\left(t_{n},x_{n}^{m}\right)\left(\log P\left(t_{n}\right)+\log P\left(x_{n}\rvert t_{n}\right)+\log P\left(y_{n}\rvert x_{n}\right)\right)dx_{n}^{m}dt_{n}\nonumber \\
 & =-\frac{\left(r+2d\right)}{2}\log\left(2\pi\right)-\frac{1}{2}\mathbb{E}_{q_{n}^{(k)}}\left[t_{n}^{T}t_{n}\right]-\frac{d}{2}\log\left(\sigma^{2}\right)\nonumber \\
 & -\frac{1}{2\sigma^{2}}\mathbb{E}_{q_{n}^{(k)}}\left[\left(x_{n}-\Lambda t_{n}-\mu_{x}\right)^{T}\left(x_{n}-\Lambda t_{n}-\mu_{x}\right)\right]-\frac{1}{2}\sum_{i=1}^{d}\log\left(\bar{x}_{i}\eta^{2}\right)\nonumber \\
 & -\frac{1}{2}\mathbb{E}_{q_{n}^{(k)}}\left[\left(y_{n}-px_{n}\right)^{T}\left[{\rm diag}\left(\bar{x}\eta^{2}\right)\right]^{-1}\left(y_{n}-px_{n}\right)\right].
\end{align}

\subsubsection{M-step}

In the M-step, considering all the available loop detector and probe
vehicle data, we maximize the sum of the expected complete log-likelihood
function in terms of the parameters $\theta$, which is 
\begin{equation}
Q\left(\theta;\theta^{(k)}\right)=\sum_{n=1}^{N}\mathbb{E}_{q_{n}^{(k)}}\left[\log\mathcal{L}\left(\theta;x_{n},y_{n},t_{n}\right)\right].
\end{equation}
The solutions to the optimization problem yield the update rules of
the parameters, which are 
\begin{equation}
\mu_{x}^{\left(k+1\right)}=\frac{1}{N}\sum_{n=1}^{N}\left(\mathbb{E}_{q_{n}^{(k)}}\left[x_{n}\right]-\Lambda^{\left(k\right)}\mathbb{E}_{q_{n}^{(k)}}\left[t_{n}\right]\right),\label{eq:update-rule-start}
\end{equation}
\begin{equation}
\Lambda^{\left(k+1\right)}=\left(\sum_{n=1}^{N}\left(\mathbb{E}_{q_{n}^{(k)}}\left[x_{n}t_{n}^{T}\right]-\mu_{x}^{\left(k\right)}\mathbb{E}_{q_{n}^{(k)}}\left[t_{n}\right]^{T}\right)\right)\left(\sum_{n=1}^{N}\mathbb{E}_{q_{n}^{(k)}}\left[t_{n}t_{n}^{T}\right]\right)^{-1},
\end{equation}
\begin{align}
\left(\sigma^{2}\right)^{\left(k+1\right)} & =\frac{1}{Nd}\sum_{n=1}^{N}\left({\rm tr}\left(\mathbb{E}_{q_{n}^{(k)}}\left[x_{n}x_{n}^{T}\right]\right)+\left(\mu_{x}^{\left(k\right)}\right)^{T}\mu_{x}^{\left(k\right)}+{\rm tr}\left(\left(\Lambda^{\left(k\right)}\right)^{T}\Lambda^{\left(k\right)}\mathbb{E}_{q_{n}^{(k)}}\left[t_{n}t_{n}^{T}\right]\right)\right.\nonumber \\
 & \left.-2\left(\mu_{x}^{\left(k\right)}\right)^{T}\mathbb{E}_{q_{n}^{(k)}}\left[x_{n}\right]-2{\rm tr}\left(\Lambda^{\left(k\right)}\mathbb{E}_{q_{n}^{(k)}}\left[x_{n}t_{n}^{T}\right]^{T}\right)+2\left(\mu_{x}^{\left(k\right)}\right)^{T}\Lambda^{\left(k\right)}\mathbb{E}_{q_{n}^{(k)}}\left[t_{n}\right]\right),
\end{align}
\begin{equation}
p^{(k+1)}=\left(\sum_{n=1}^{N}y_{n}^{T}{\rm diag}\left(\bar{x}\left(\eta^{2}\right){}^{\left(k\right)}\right)^{-1}\mathbb{E}_{q_{n}^{(k)}}\left[x_{n}\right]\right)\left(\sum_{n=1}^{N}{\rm tr}\left(\mathbb{E}_{q_{n}^{(k)}}\left[x_{n}x_{n}^{T}\right]{\rm diag}\left(\bar{x}\left(\eta^{2}\right){}^{\left(k\right)}\right)^{-1}\right)\right)^{-1},
\end{equation}
\begin{equation}
\left(\eta^{2}\right)^{(k+1)}=\frac{1}{Nd}\sum_{n=1}^{N}\sum_{i=1}^{d}\frac{1}{\bar{x}_{i}}\left(y_{ni}^{2}-2p^{\left(k\right)}y_{ni}\mathbb{E}_{q_{n}^{(k)}}\left[x_{n}\right]_{i}+\left(p^{\left(k\right)}\right)^{2}\mathbb{E}_{q_{n}^{(k)}}\left[x_{n}x_{n}^{T}\right]_{ii}\right).\label{eq:update-rule-end}
\end{equation}
The solutions are concisely expressed in terms of five expectations
derived from Eq. (\ref{eq:derive5expectations}). The detailed solving
process and the expressions of the expectations are in Appendix A.

\subsection{Estimating the unknown traffic volumes}

Performing the update rules introduced above iteratively leads to
the convergence of the estimated $\theta$ \citep{dempster1977maximum}.
With the estimated parameters of the PPCA-DF model, the posterior
predictive distribution of the unknown data is a Gaussian distribution
given by
\begin{equation}
x_{n}^{m}\rvert x_{n}^{o},y_{n}\sim\mathcal{N}\left(\mu_{x_{n}^{m}\rvert x_{n}^{o},y_{n}},\Sigma_{x_{n}^{m}\rvert x_{n}^{o},y_{n}}\right),
\end{equation}
where
\begin{align}
\mu_{x_{n}^{m}\rvert x_{n}^{o},y_{n}} & =\mu_{x_{n}^{m}}+\left[\begin{array}{cc}
\Sigma_{x_{n}^{m}x_{n}^{o}} & \Sigma_{x_{n}^{m}y}\end{array}\right]\left[\begin{array}{cc}
\Sigma_{x_{n}^{o}x_{n}^{o}} & \Sigma_{x_{n}^{o}y_{n}}\\
\Sigma_{y_{n}x_{n}^{o}} & \Sigma_{y_{n}y_{n}}
\end{array}\right]^{-1}\left(\left[\begin{array}{c}
x_{n}^{o}\\
y_{n}
\end{array}\right]-\left[\begin{array}{c}
\mu_{x_{n}^{o}}\\
\mu_{y}
\end{array}\right]\right),\\
\Sigma_{x_{n}^{m}\rvert x_{n}^{o},y_{n}} & =\Sigma_{x_{n}^{m}x_{n}^{m}}-\left[\begin{array}{cc}
\Sigma_{x_{n}^{m}x_{n}^{o}} & \Sigma_{x_{n}^{m}y_{n}}\end{array}\right]\left[\begin{array}{cc}
\Sigma_{x_{n}^{o}x_{n}^{o}} & \Sigma_{x_{n}^{o}y_{n}}\\
\Sigma_{y_{n}x_{n}^{o}} & \Sigma_{y_{n}y_{n}}
\end{array}\right]^{-1}\left[\begin{array}{cc}
\Sigma_{x_{n}^{m}x_{n}^{o}} & \Sigma_{x_{n}^{m}y_{n}}\end{array}\right]^{T}.
\end{align}
Therefore, we can estimate the unknown traffic volumes in the $n$th
column of $X$ by the mean $\mu_{x_{n}^{m}\rvert x_{n}^{o},y_{n}}$. 

\section{Case studies \label{sec:Case-studies}}

In this section, we first introduce the dataset we use for validation.
Then, we validate the proposed method in different scenarios and compare
its performance with the baseline methods.

\subsection{Ground-truth dataset}

\begin{figure}[H]
\begin{centering}
\includegraphics[width=1\textwidth]{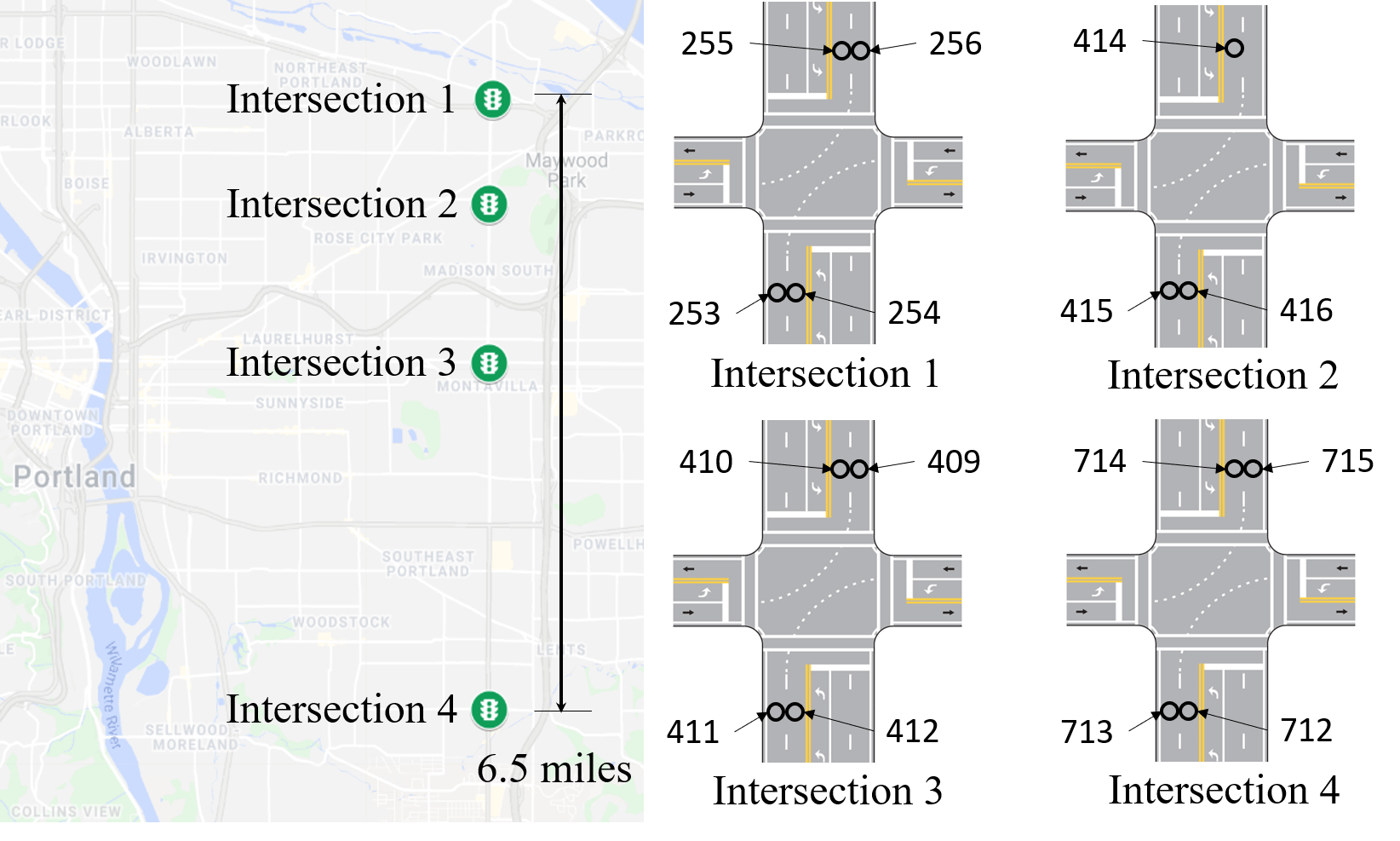}
\par\end{centering}
\caption{Ground-truth dataset locations. \label{fig:82nd-Ave}}
\end{figure}

\begin{figure}[H]
\begin{centering}
{\small{}\includegraphics[width=1\textwidth]{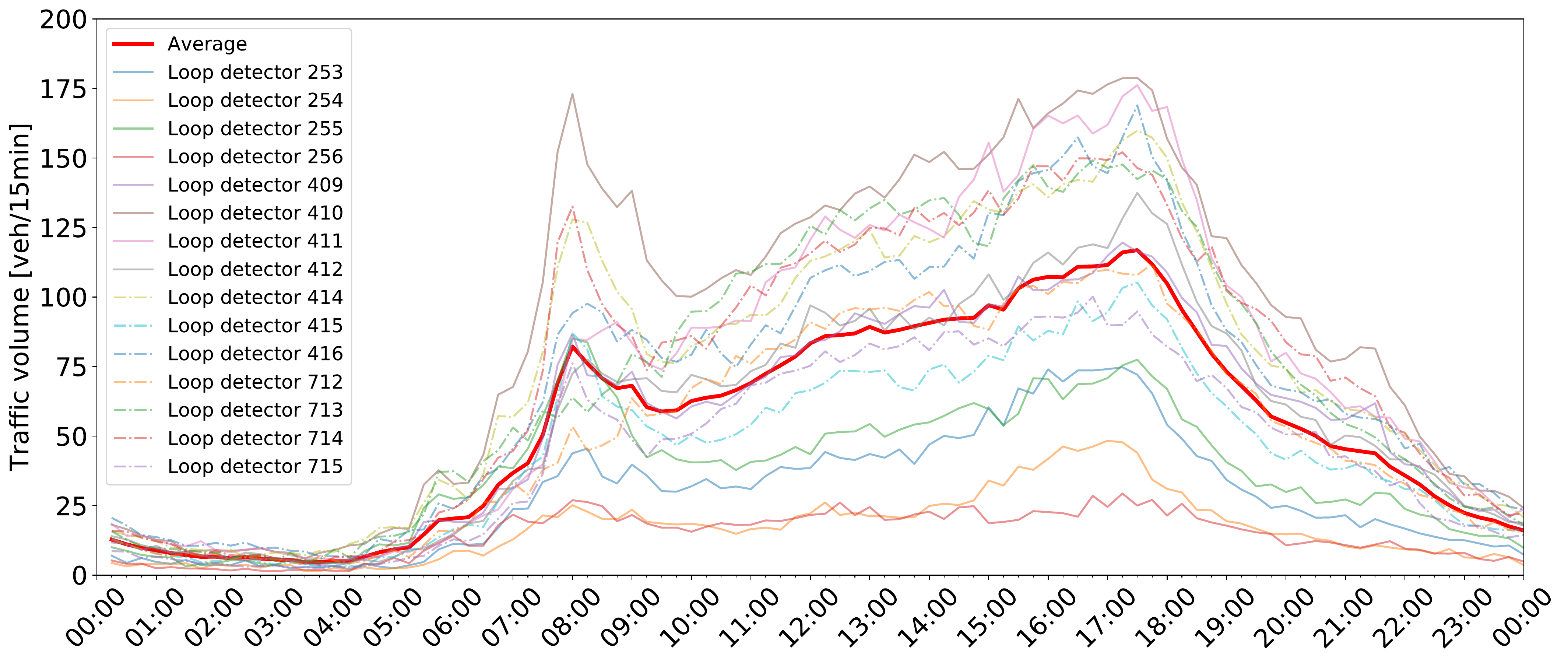}}{\small\par}
\par\end{centering}
\caption{The average traffic volumes over the 15 workdays at different locations
and TODs. \label{fig:Average-traffic-volume}}
\end{figure}

To examine the performance of the proposed PPCA-DF model in both the
missing data scenario and the low coverage scenario, we conduct experiments
using a real-world traffic volume dataset. The ground-truth dataset
used here is the PORTAL Arterial Data (https://portal.its.pdx.edu/fhwa)
collected from the loop detectors on 82nd Avenue in Portland, Oregon.
The IDs of the specific 15 loop detectors we use are 253, 254, 255,
256, 409, 410, 411, 412, 414, 415, 416, 712, 713, 714, and 715. The
locations of the loop detectors along 82nd Avenue are shown in Figure
\ref{fig:82nd-Ave}. Because of the data availability, we choose four intersections (Intersection 1 to Intersection 4 shown on Figure \ref{fig:82nd-Ave}) spanning roughly 6.5 miles apart. We use the data of 15 workdays from October
21 to November 10, 2011. We aggregate the data to 15-min intervals
in the preprocessing stage. Figure \ref{fig:Average-traffic-volume}
shows the average traffic volumes in different TODs over the 15 workdays
collected by the 15 loop detectors. In general, the traffic volumes
at different locations fluctuate in a similar trend over time, which
implies a strong correlation.

\subsection{Experimental settings}

\subsubsection{Probe vehicle data}

Since we do not have access to the real-world probe vehicle data collected
from the studied locations at the same time, for validation purposes,
we simulate the probe vehicle data by randomly sampling from the whole-population
traffic volume data. For each vehicle recorded by the loop detectors,
we randomly determine if it is a probe vehicle or a regular vehicle
according to the penetration rate $p$. After this step, we obtain
the simulated probe vehicle data, which are further converted into
the probe vehicle traffic volume matrix $Y$.

\subsubsection{Traffic volume data with missing entries}

We simulate two missing data patterns to characterize the two different
scenarios of our interest, i.e., the missing data scenario and the
low coverage scenario. For the missing data scenario, given a missing
ratio, we perform a Bernoulli trial to decide if each entry in the
ground-truth traffic volume matrix is missing or not. If the random
sampling process shows the entry is missing, we will hide the value
of the entry. This process simulates the loop detector malfunction
situation. For the low coverage scenario, we randomly remove several
rows of the ground-truth traffic volume matrix. This process simulates
the situation where some of the studied locations are not covered
by loop detectors. After this step, we obtain the simulated loop detector
traffic volume matrix $X$. In this case study, for each TOD, the
size of $X$ and $Y$ is $15\times15$, as there are 15 loop detectors
and 15 days. The loop detector traffic volume matrix $X$ (with missing
entries or rows), together with the probe vehicle traffic volume matrix
$Y$, will serve as the input to our proposed model.

\subsection{Measure of accuracy}

We evaluate the performance of the proposed method by the root mean
square error (RMSE) when reconstructing the missing entries. The traffic
volumes are reconstructed for each TOD separately. Then, we calculate
the performance measure for each TOD, or combine results for all TODs
to get the overall performance measurement.

\subsection{Results for the missing data scenario}

\begin{figure}[H]
\begin{centering}
\includegraphics[width=1\textwidth]{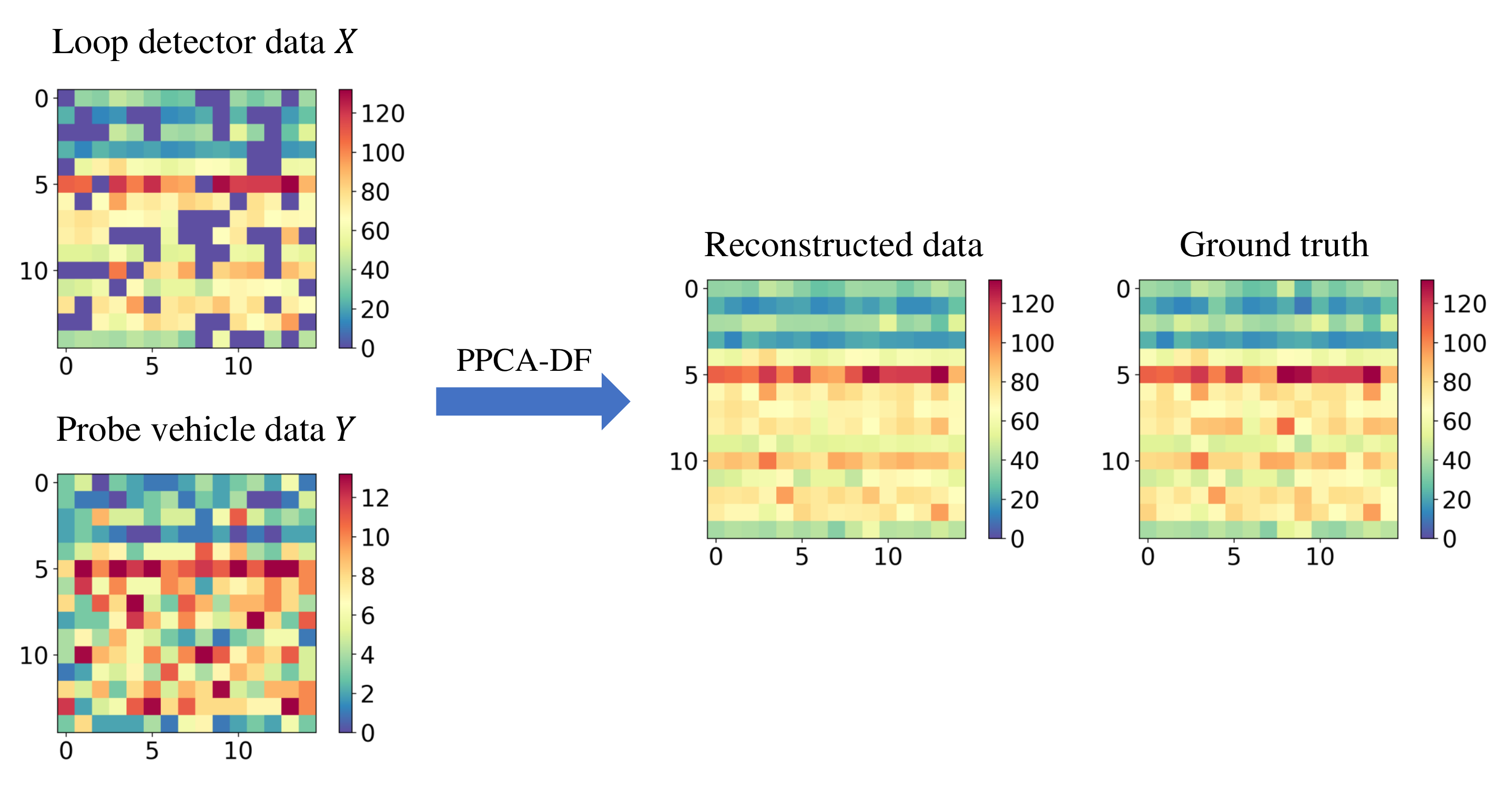}
\par\end{centering}
\caption{Traffic volume reconstruction for the missing data scenario.\label{fig:Illustration-missing-data-scenario}}
\end{figure}

Figure \ref{fig:Illustration-missing-data-scenario} illustrates the
estimation process in the missing data scenario. The input data include
the loop detector data $X$ with randomly missing entries and the
probe vehicle traffic volume matrix $Y$. The colors represent different
magnitudes of traffic volumes. The entries with traffic volumes equal
to zero correspond to those entries with missing data. Using the proposed
PPCA-DF model, we can reconstruct the missing traffic volumes.

The dimension of the projection matrix $\Lambda$ is $d\times r$,
where $r$ is the rank of the matrix. Intuitively, increasing the
rank of the projection matrix can capture more spatiotemporal correlations
in the traffic volume data. However, increasing the rank may also
increase the model complexity and result in overfitting. In the case
study, we consider the traffic volume reconstruction for 15-min intervals.
Due to the small granularity, the variance of the probe vehicle traffic
volume is relatively large. As a result, if the rank is too large,
the noise in data will be introduced to the model. This effect is
particularly critical during the night time, when the ground-truth
traffic volume is very small and the number of observed probe vehicles
fluctuate significantly. Therefore, after doing the cross-validation
process, we set $r=1$. In other settings, for example, when we reconstruct
60-min traffic volumes, increasing the rank might give rise to better
performance.

\subsubsection{The impact of missing ratios and penetration rates}

In this section, we examine the impact of missing ratios and penetration
rates on the PPCA-DF model in the missing data scenario. We enumerate
the missing ratio from 5\% to 95\%, with a step size of 5\%. At the
same time, we test the method under different penetration rates, including
5\%, 10\%, 20\%, and 50\%. For each test, we run 30 independent experiments
and then calculate the average performance measurement.

\begin{figure}[H]
\begin{centering}
\includegraphics[width=0.5\textwidth]{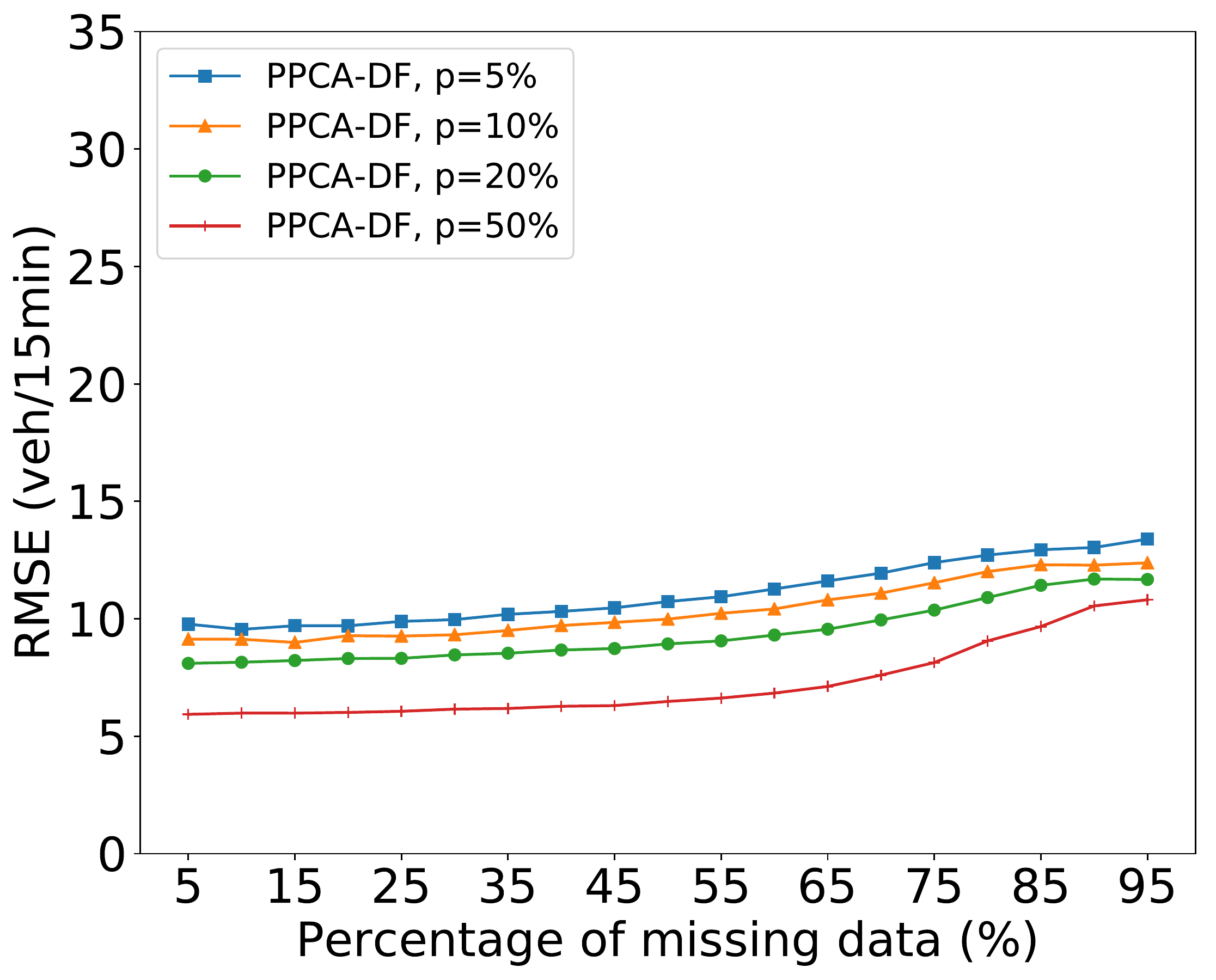}
\par\end{centering}
\caption{\label{fig:randon-prr}Performance of the proposed method under different
missing ratios and different penetration rates in the missing data
scenario.}
\end{figure}

Figure \ref{fig:randon-prr} shows the results. In general, the estimation
accuracy decreases as the missing ratio increases. It is because when
the missing ratio is low, non-missing entries can provide abundant
information. By contrast, when the missing ratio is high, the number
of remaining entries is very limited, which makes it challenging to
estimate the unknown traffic volumes. Even though, the proposed model
can still give an accurate estimation when the missing ratio is higher
than 80\%, which validates the robustness of our approach. Overall,
the proposed method is not very sensitive to the missing ratio. It
is due to the benefits of fusing the loop detector data and probe
vehicle data. Probe vehicles cover a broad range of locations and
provide a partial observation of the traffic volumes, whereas loop
detectors measure the traffic volumes at several discrete locations.
Even though the missing ratio is high and many entries are missing,
we can still estimate the corresponding traffic volumes by exploiting
the spatiotemporal correlations contained in the two data sources.

The probe vehicle penetration rate is another critical parameter that
influences the performance of the method. It determines the magnitude
of the probe vehicle traffic volumes. With a higher penetration rate,
the spatiotemporal correlation of the traffic volume data can be better
retained in the probe vehicle data. Even though, the results in Figure
\ref{fig:randon-prr} show that the proposed method can already reconstruct
traffic volumes accurately when the penetration rate is only 10\%.
These results indicate the feasibility of practical applications of
our approach, as some studies have already shown that the probe vehicle
penetration rate could reach 10\% in some places \citep{zheng2017estimating,zhao2019volume,zhao2019various}.
For illustration purposes, we will use 10\% as the penetration rate
in the following experiments.

\subsubsection{Comparison with existing methods}

We compare the proposed PPCA-DF model with two baseline methods. The
first one is the direct scaling method, which reconstructs the unknown
traffic volumes by scaling up the traffic volumes of the probe vehicles
using the penetration rate directly \citep{wong2019estimation}. The
second method is the PPCA method used by \citet{qu2009ppca} and \citet{li2013efficient},
which captures the low-rank structure by solving an MLE problem. For
the PPCA method, it only uses the loop detector data to construct
the low-rank structure of the traffic volume data and then uses it
for missing value imputation. It also needs to determine the dimension
of the projection matrix. We choose the dimension that yields the
best performance, which is also $r=1$ for the 15-min interval case.

\begin{figure}[H]
\begin{centering}
\includegraphics[width=0.5\textwidth]{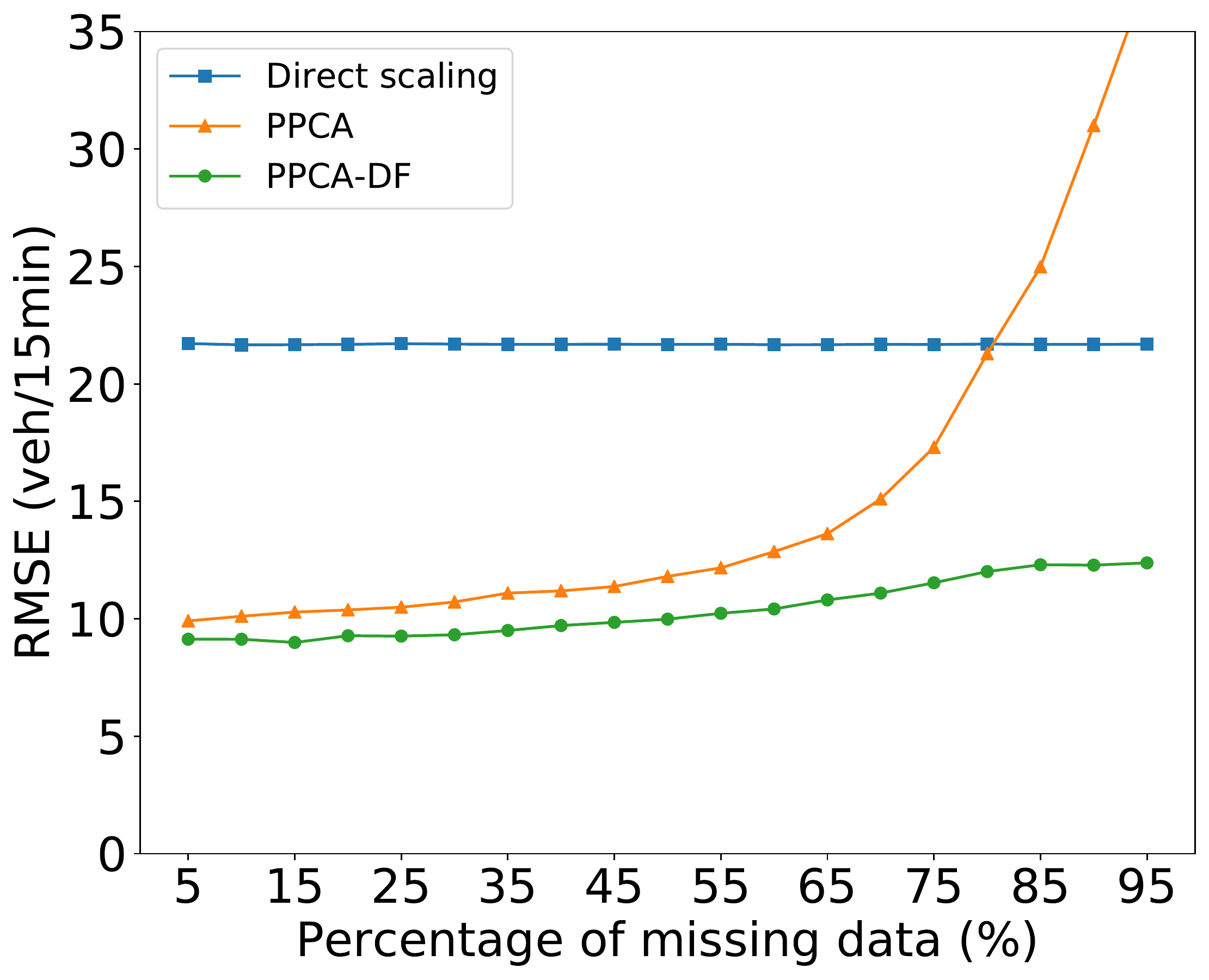}
\par\end{centering}
\caption{\label{fig:randon-compare}Comparison of different methods for the
missing data scenario.}
\end{figure}

The comparison results are shown in Figure \ref{fig:randon-compare}.
As the results suggest, the proposed method consistently outperforms
the two baseline methods under different missing ratios. For the direct
scaling method, it only utilizes the probe vehicle data, so the estimation
accuracy does not depend on the missing ratio. Scaling up the probe
vehicle traffic volume directly will amplify its variance, especially
when the penetration rate is not high enough or when the magnitude
of traffic volume is low (e.g., at the night time). However, the direct
scaling method does not use any information from other locations and
time slots to reduce the variance. Therefore, the RMSE is relatively
large. The proposed method also yields better performance than the
PPCA method for all missing ratios. Especially when the missing ratio
is high, the PPCA baseline method cannot reconstruct the missing values
accurately with limited information. The reason is that, when the
missing ratio is high, many entries in the loop detector data are
missing, which severely undermines the spatiotemporal correlation
the traffic volume data should have. The PPCA-DF model proposed in
this paper, however, tries to find the low-rank representation of
the traffic volumes by combining both of the data sources, which gives
rise to better estimation accuracy. The results imply that the probe
vehicle data is an appropriate data source for finding the embedded
spatiotemporal correlations. It also validates the idea that incorporating
probe vehicle data can provide a robust approach to the reconstruction
of traffic volumes.

\begin{figure}[H]
\begin{centering}
\subfloat[\label{fig:randon-compare-TOD-RMSE}]{\begin{centering}
\includegraphics[width=1\textwidth]{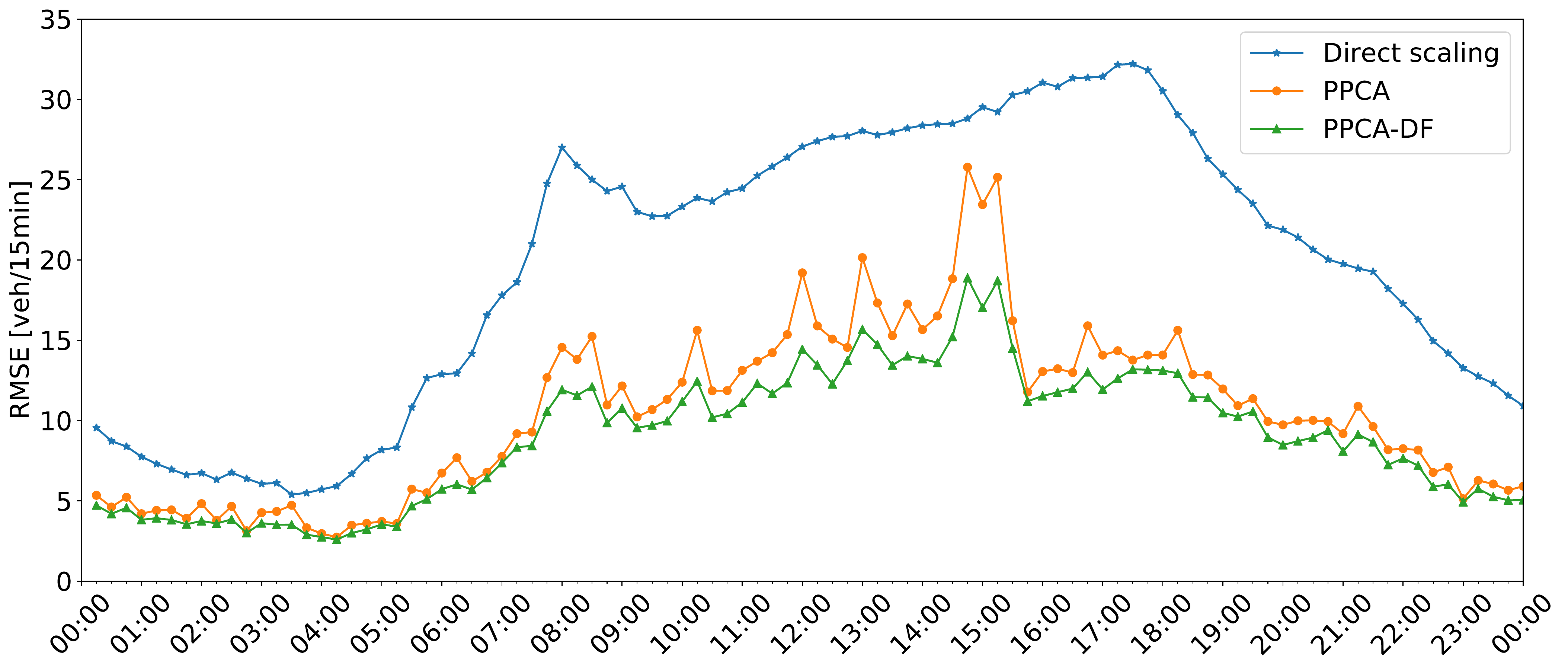}
\par\end{centering}
}
\par\end{centering}
\begin{centering}
\subfloat[\label{fig:randon-compare-TOD-MAPE}]{\begin{centering}
\includegraphics[width=1\textwidth]{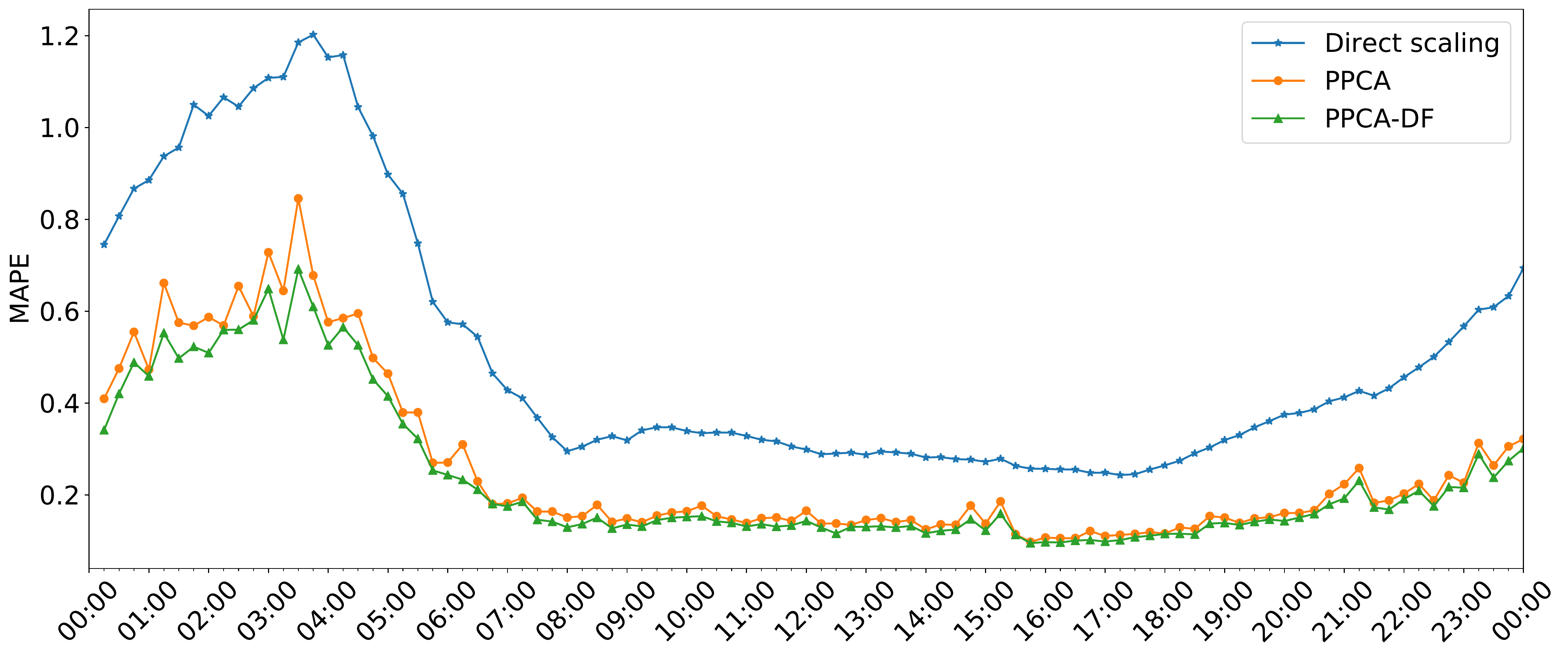}
\par\end{centering}
}
\par\end{centering}
\caption{\label{fig:randon-compare-TOD}Estimation accuracy of different methods
in different TODs for the missing data scenario measured by: (a) RMSE
and (b) MAPE.}
\end{figure}

We further examine the estimation accuracy of the methods in different
TODs. Figure \ref{fig:randon-compare-TOD-RMSE} shows the estimation
results of different methods. The figure corresponds to the scenario
where the percentage of missing data is 50\%. The proposed method
outperforms the direct scaling and PPCA baseline methods in almost
all the TODs. The RMSE is smaller in the night time compared to the
day time, because the ground-truth traffic volumes are much smaller
in the night time, as shown in Figure \ref{fig:Average-traffic-volume}.
The mean absolute percentage error (MAPE) is actually smaller in the
day time as shown in Figure \ref{fig:randon-compare-TOD-MAPE}, due
to the larger sample size of probe vehicle data and high spatiotemporal
correlation. The results show that the MAPE is around 13\% in the
day time (9:00-17:00), which suggests good estimation accuracy for
15-min traffic volumes.

\subsection{Results for the low coverage scenario}

\begin{figure}[H]
\begin{centering}
\includegraphics[width=1\textwidth]{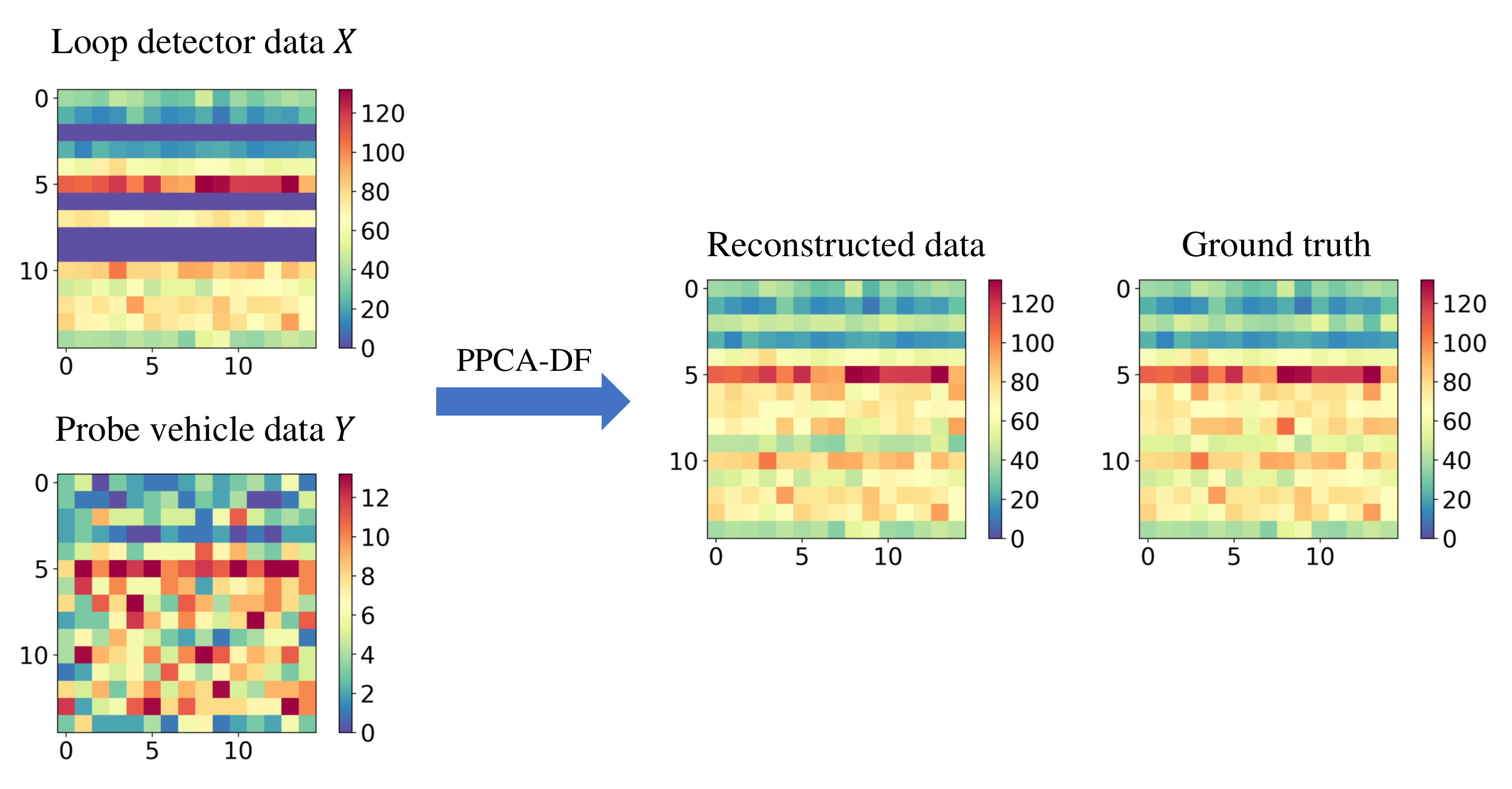}
\par\end{centering}
\caption{Traffic volume reconstruction for the low coverage scenario. \label{fig:Illustration-low-coverage}}
\end{figure}

The low coverage scenario is more challenging. In this case, not all
the locations we study are covered by loop detectors. In other words,
several rows of the traffic volume matrix $X$ can be missing. Figure
\ref{fig:Illustration-low-coverage} illustrates the whole process
of traffic volume reconstruction for the low coverage scenario. Similar
to the missing data scenario, the input data include the loop detector
data $X$ with some missing rows and the probe vehicle data $Y$.
The unknown traffic volumes can be reconstructed using the proposed
PPCA-DF model.

\subsubsection{The impact of missing ratios and penetration rates}

\begin{figure}[H]
\begin{centering}
\includegraphics[width=0.5\textwidth]{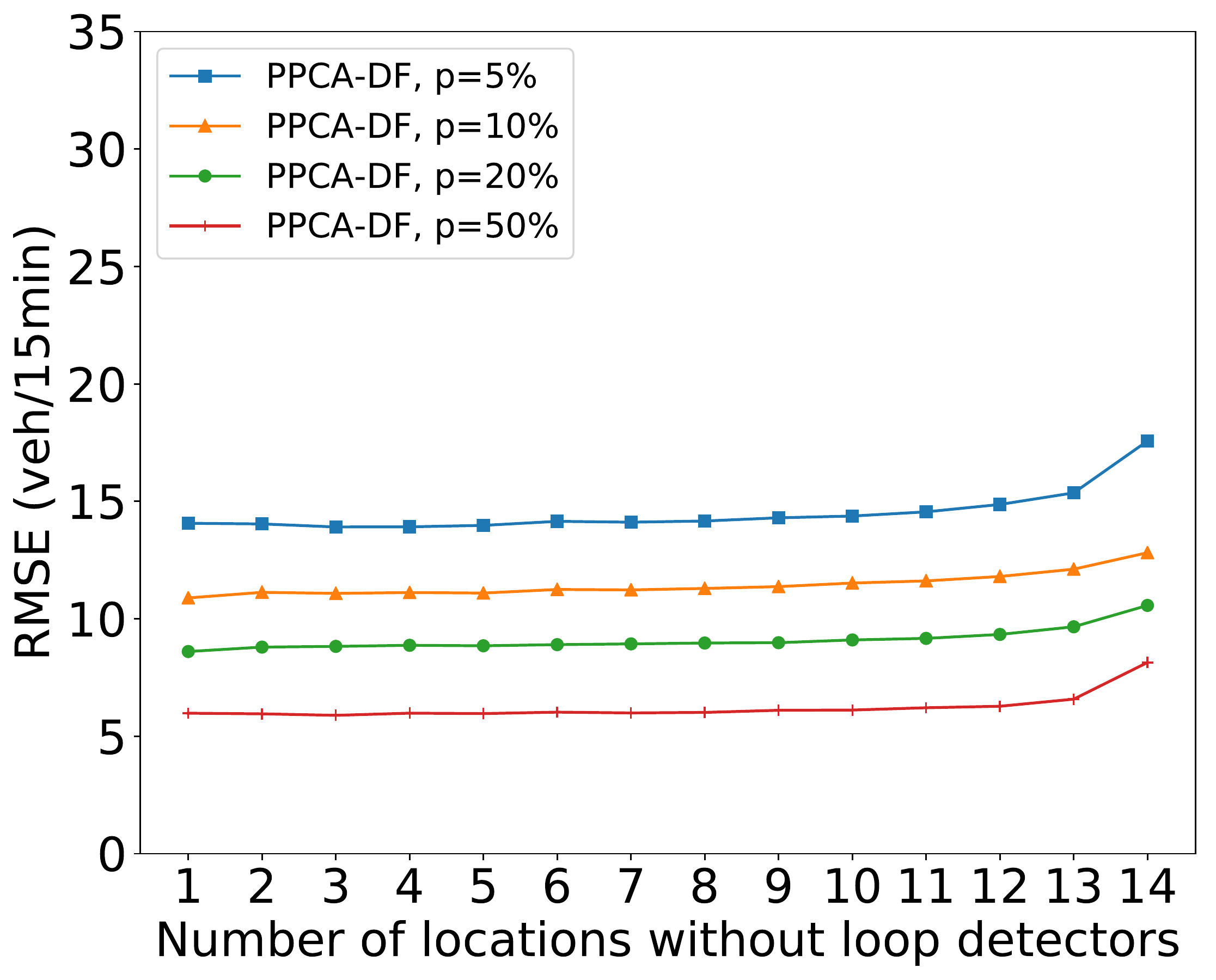}
\par\end{centering}
\caption{\label{fig:low-coverage-prr}Performance of the proposed method under
different missing ratios and different penetration rates in the low
coverage scenario.}
\end{figure}

We first evaluate the performance of the proposed model under different
missing ratios and different levels of probe vehicle market penetration.
Figure \ref{fig:low-coverage-prr} shows the results. Similar to the
missing data scenario, generally speaking, a lower missing ratio or
a higher penetration rate leads to better estimation accuracy. The
results suggest that a 10\% penetration rate can still enable the
proposed method to reconstruct traffic volumes accurately, even when
multiple locations are not covered with loop detectors. Again, considering
the current situations of probe vehicle deployment, we will still
use the 10\% penetration rate in the following experiments.

From the results, we can also find that when there are more locations
not covered by loop detectors, the estimation accuracy does not decrease
drastically, which again demonstrates the robustness of the proposed
method. It is still because of the benefits of involving the probe
vehicle data source. Although many locations are not covered by loop
detectors, the spatiotemporal correlation can be captured by probe
vehicle data which covers all locations. Therefore, by leveraging
data fusion, we can still accurately estimate the corresponding traffic
volumes. It implies that the probe vehicle data is a valuable complement
to the traditional loop detector data, and it can help us better understand
the traffic flows at the network level.

\subsubsection{Comparison with existing methods}

\begin{figure}[H]
\begin{centering}
\includegraphics[width=0.5\textwidth]{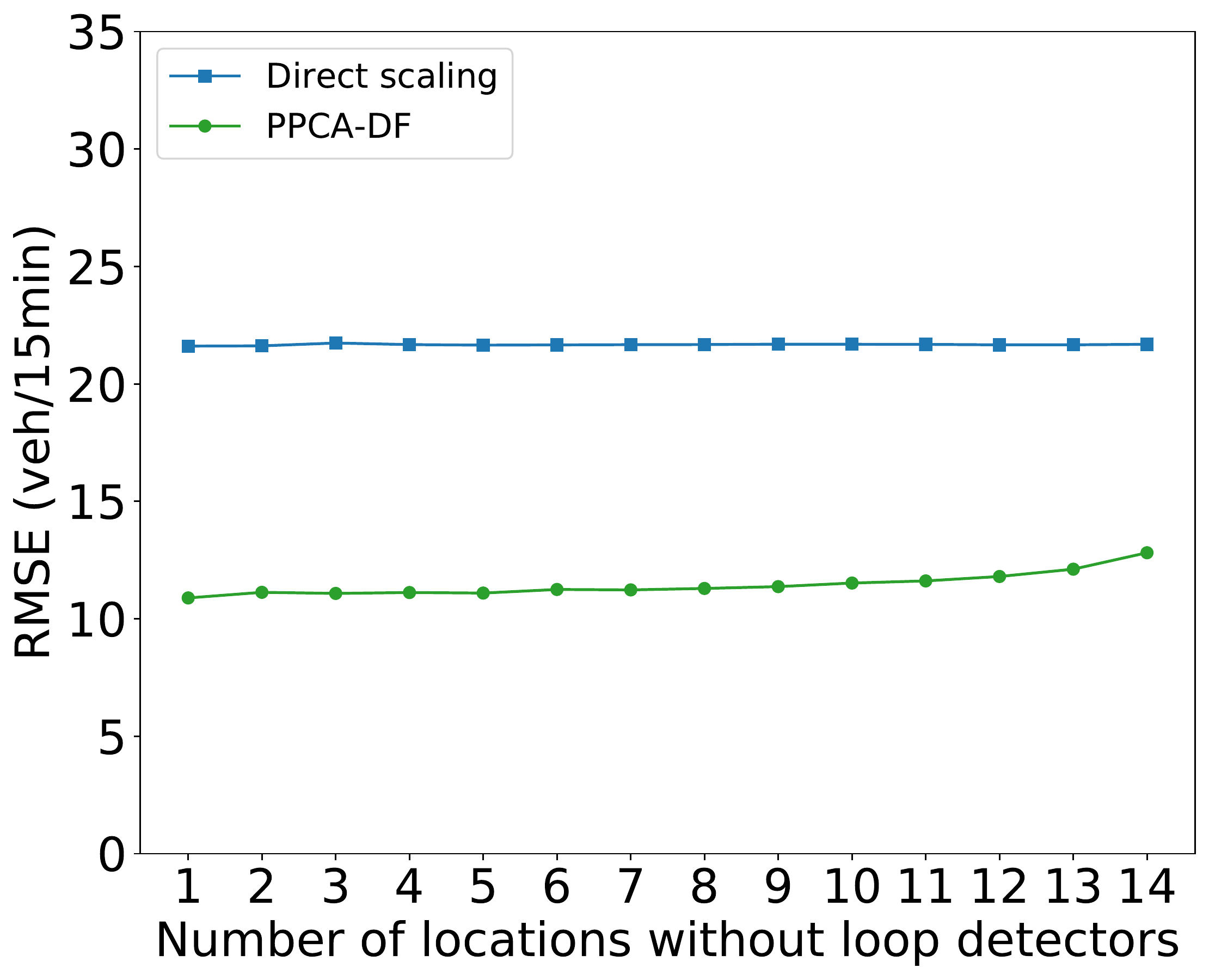}
\par\end{centering}
\caption{\label{fig:low-coverage-compare}Comparison of different methods for
the low coverage scenario.}
\end{figure}

Since the PPCA method used in \citet{qu2009ppca} and \citet{li2013efficient}
cannot deal with the low coverage scenario, we only compare the proposed
PPCA-DF model with the direct scaling method. As shown in Figure \ref{fig:low-coverage-compare},
the proposed method performs significantly better than the baseline
method. It is because the proposed method considers the spatiotemporal
correlation in the traffic volume data, whereas the direct scaling
method considers each location and each time slot independently.

\begin{figure}[H]
\begin{centering}
\subfloat[\label{fig:low-coverage-compare-TOD-RMSE}]{\begin{centering}
\includegraphics[width=1\textwidth]{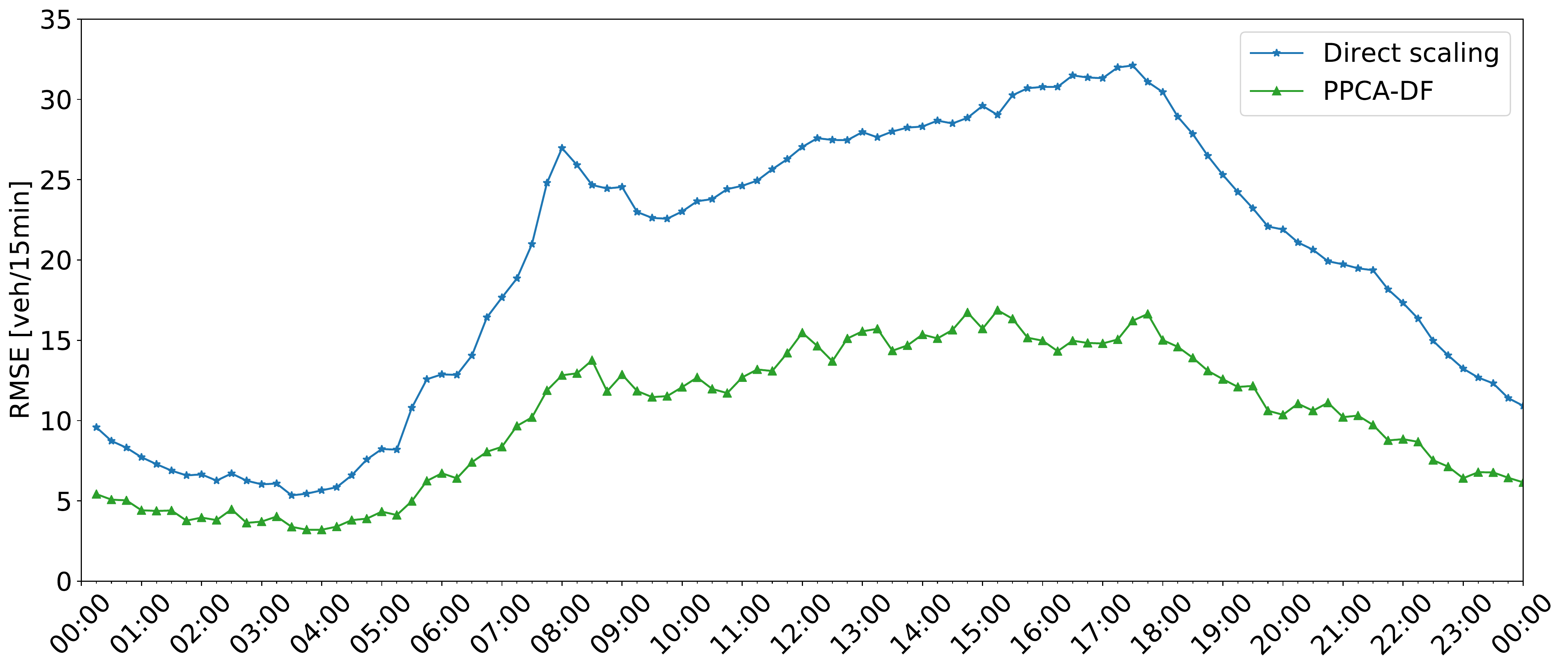}
\par\end{centering}
}
\par\end{centering}
\begin{centering}
\subfloat[\label{fig:low-coverage-compare-TOD-MAPE}]{\begin{centering}
\includegraphics[width=1\textwidth]{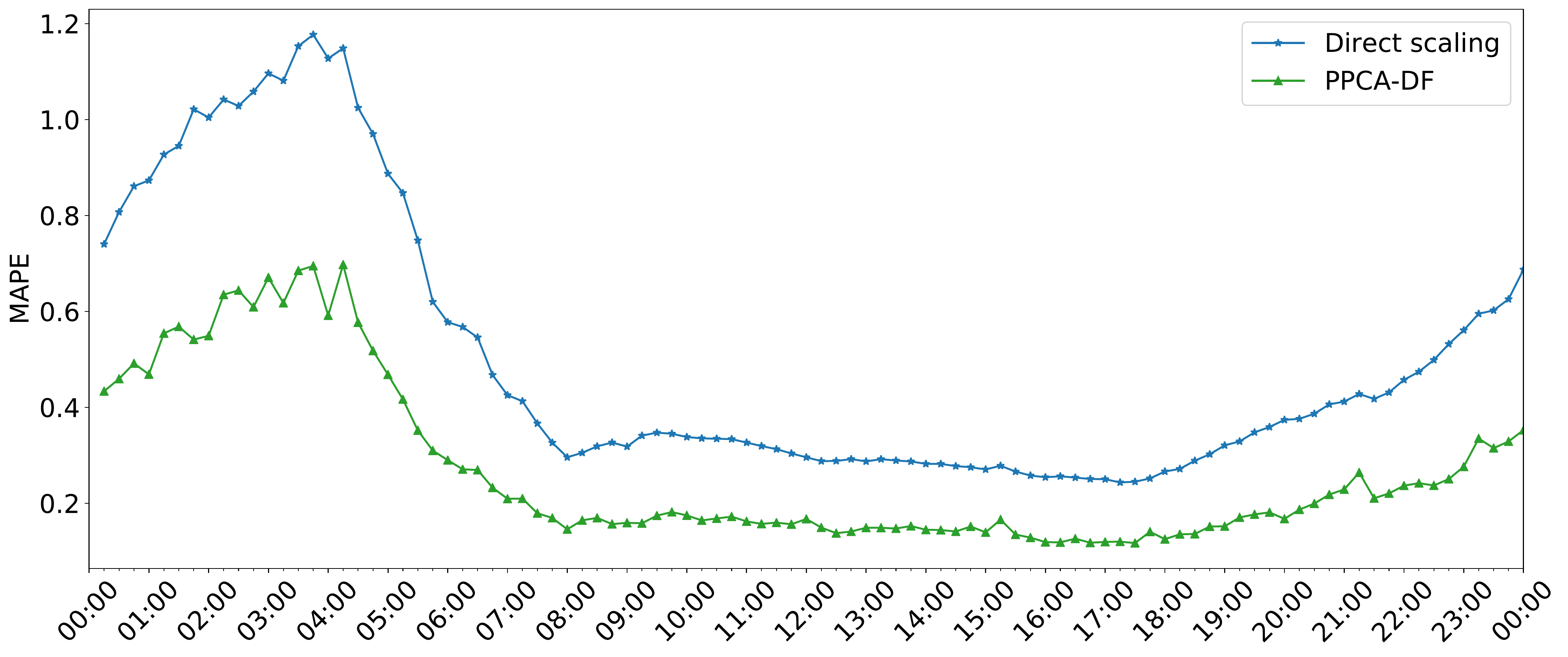}
\par\end{centering}
}
\par\end{centering}
\caption{\label{fig:low-coverage-compare-TOD}Estimation accuracy of different
methods in different TODs for the low coverage scenario measured by:
(a) RMSE and (b) MAPE.}
\end{figure}

Figure \ref{fig:low-coverage-compare-TOD} shows the estimation results
of different methods in different TODs. The figure corresponds to
the scenario when seven randomly chosen locations are not covered
by loop detectors. The proposed method outperforms the direct scaling
method in all the TODs. The MAPE is around 15\% in the day time (9:00-17:00),
which validates the performance of the proposed method in the low
coverage scenario.

\section{Conclusions \label{sec:Conclusions}}

In this paper, we first propose a general probabilistic framework
for traffic state estimation problems. Based on the framework, we
propose a data fusion model PPCA-DF, which can reconstruct the unknown
traffic volumes by exploiting the spatiotemporal correlations of traffic
volumes. The proposed PPCA-DF model considers both loop detector data
and probe vehicle data when inferring the low-rank structure using
maximum likelihood estimation. The PPCA-DF model can handle two critical
and frequently occurring scenarios. The first scenario is the missing
data scenario, where some traffic volume data are missing due to loop
detector malfunction or communication failure. This problem has long
been commonly recognized as a shortcoming of loop detectors. The second
scenario is the low coverage scenario, where no loop detectors are
installed at some locations of our interest. This scenario also frequently
occurs, since fixed-location sensors usually can only cover a small
subset of links in the transportation network due to the high installation
and maintenance costs. Whereas most of the existing literature focuses
on the first scenario and cannot handle the second scenario, the proposed
PPCA-DF model can be applied to both scenarios.

We examine the performance of the proposed method using a real-world
loop detector dataset collected from Portland, Oregon. The results
show that the PPCA-DF model can achieve good performance when dealing
with both of the two scenarios. It can outperform the existing methods
even when the penetration rate of probe vehicles is only 10\%. What
is more, the proposed model is not sensitive to the missing ratio
and can give accurate estimation results when the missing ratio is
higher than 80\% or when most locations have no loop detectors installed.
These results validate the effectiveness and robustness of the proposed
model and also imply its potential for practical applications.

The current work can be extended in a few directions by future research.
For instance, firstly, we only count the number of probe vehicles
passing by a certain location and use the aggregated information in
the model. However, the probe vehicle data include the complete trajectories
of the vehicles, which contain rich information. The estimation accuracy
could be further improved by considering more detailed information
encoded in the trajectory, such as the arrival time of each probe
vehicle. Secondly, we assume that the penetration rate of probe vehicles
at each of the studied locations is the same in the same TOD. The
assumption can be relaxed by considering the penetration rate as a
random variable following a certain distribution. It will give more
flexibility to the model and can therefore capture the daily variance
of the penetration rate. Thirdly, in this study, we assume the observed
loop detector volume is the ground-truth traffic volume. However,
in a real-world situation, measurement errors for the loop detector
do exist, and the errors should also be considered.

\section*{Appendix A \label{sec:Appendix-A-EM-solution}}

The analytical solutions of the EM algorithm can be obtained by setting
the derivatives of $Q\left(\theta;\theta^{(k)}\right)$ to zero, i.e.,
\begin{equation}
\frac{\partial Q\left(\theta\mid\theta^{(k)}\right)}{\partial\mu_{x}}=\sum_{n=1}^{N}\frac{1}{\left(\sigma^{2}\right)^{\left(k\right)}}\left(\mathbb{E}_{q_{n}^{(k)}}\left[x_{n}\right]-\Lambda^{\left(k\right)}\mathbb{E}_{q_{n}^{(k)}}\left[t_{n}\right]-\mu_{x}\right)=0,
\end{equation}
\begin{equation}
\frac{\partial Q\left(\theta\mid\theta^{(k)}\right)}{\partial\Lambda}=\sum_{n=1}^{N}\frac{1}{\left(\sigma^{2}\right)^{\left(k\right)}}\left(\mathbb{E}_{q_{n}^{(k)}}\left[\left(x_{n}-\mu_{x}^{(k)}\right)t_{n}^{T}\right]-\Lambda\mathbb{E}_{q_{n}^{(k)}}\left[t_{n}t_{n}^{T}\right]\right)=0,
\end{equation}
\begin{equation}
\frac{\partial Q\left(\theta\mid\theta^{(k)}\right)}{\partial\sigma^{2}}=\sum_{n=1}^{N}\left(\frac{d}{\sigma^{2}}-\frac{1}{\sigma^{4}}\mathbb{E}_{q_{n}^{(k)}}\left[\left(x_{n}-\Lambda^{\left(k\right)}t_{n}-\mu_{x}^{(k)}\right)^{T}\left(x_{n}-\Lambda^{\left(k\right)}t_{n}-\mu_{x}^{(k)}\right)\right]\right)=0,
\end{equation}
\begin{equation}
\frac{\partial Q\left(\theta\mid\theta^{(k)}\right)}{\partial p}=\sum_{n=1}^{N}\mathbb{E}_{q_{n}^{(k)}}\left[\left({\rm diag}\left(\bar{x}\left(\eta^{2}\right){}^{\left(k\right)}\right)^{-1}\left(y_{n}-px_{n}\right)\right)^{T}x_{n}\right]=0,
\end{equation}
\begin{equation}
\frac{\partial Q\left(\theta\mid\theta^{(k)}\right)}{\partial\eta^{2}}=\sum_{n=1}^{N}\left(\frac{d}{\eta^{2}}-\frac{1}{\eta^{4}}\mathbb{E}_{q_{n}^{(k)}}\left[\left(y_{n}-p^{\left(k\right)}x_{n}\right)^{T}{\rm diag}\left(\bar{x}\right)^{-1}\left(y_{n}-p^{\left(k\right)}x_{n}\right)\right]\right)=0.
\end{equation}
 Solving the equations above yields the update rules of the parameters,
i.e., Eqs. (\ref{eq:update-rule-start})-(\ref{eq:update-rule-end}).
The explicit expressions of the five expectations are shown as below.
\begin{equation}
\mathbb{E}_{q_{n}^{(k)}}\left[x_{n}\right]=\left[\begin{array}{c}
\mu_{x_{n}^{m}\rvert x_{n}^{o},y_{n}}^{(k)}\\
x_{n}^{o}
\end{array}\right],
\end{equation}
\begin{equation}
\mathbb{E}_{q_{n}^{(k)}}\left[t_{n}\right]=\mu_{t_{n}\rvert x_{n}^{o},y_{n}}^{(k)},
\end{equation}
\begin{equation}
\mathbb{E}_{q_{n}^{(k)}}\left[t_{n}t_{n}^{T}\right]=\Sigma_{t_{n}\rvert x_{n}^{o},y_{n}}^{(k)}+\mu_{t_{n}\rvert x_{n}^{o},y_{n}}^{(k)}\left(\mu_{t_{n}\rvert x_{n}^{o},y_{n}}^{(k)}\right)^{T},
\end{equation}
\begin{equation}
\mathbb{E}_{q_{n}^{(k)}}\left[x_{n}x_{n}^{T}\right]=\left[\begin{array}{cc}
\Sigma_{x_{n}^{m}\rvert x_{n}^{o},y_{n}}^{\left(k\right)}+\mu_{x_{n}^{m}\rvert x_{n}^{o},y_{n}}^{(k)}\left(\mu_{x_{n}^{m}\rvert x_{n}^{o},y_{n}}^{(k)}\right)^{T} & \mu_{x_{n}^{m}\rvert x_{n}^{o},y_{n}}^{(k)}\left(x_{n}^{o}\right)^{T}\\
x_{n}^{o}\left(\mu_{x_{n}^{m}\rvert x_{n}^{o},y_{n}}^{(k)}\right)^{T} & x_{n}^{o}\left(x_{n}^{o}\right)^{T}
\end{array}\right],
\end{equation}
\begin{equation}
\mathbb{E}_{q_{n}^{(k)}}\left[x_{n}t_{n}^{T}\right]=\left[\begin{array}{c}
\Sigma_{x_{n}^{m}t_{n}\rvert x_{n}^{o},y_{n}}^{(k)}+\mu_{x_{n}^{m}\rvert x_{n}^{o},y_{n}}^{(k)}\left(\mu_{t_{n}\rvert x_{n}^{o},y_{n}}^{(k)}\right)^{T}\\
x_{n}^{o}\left(\mu_{t_{n}\rvert x_{n}^{o},y_{n}}^{(k)}\right)^{T}
\end{array}\right].
\end{equation}

\section*{Acknowledgment}

The authors would like to thank the US Department of Transportation
(USDOT) Region 5 University Transportation Center: Center for Connected
and Automated Transportation (CCAT) of the University of Michigan
for funding the research. The views presented in this paper are those
of the authors alone. 

\section*{CRediT author statement}

\textbf{Xintao Yan: }Conceptualization, Methodology, Software, Data
curation, Writing-original draft, Writing-review and editing.\textbf{
Yan Zhao: }Conceptualization, Methodology, Software, Data curation,
Writing-original draft, Writing-review and editing\textbf{. Henry
X. Liu: }Conceptualization, Writing-review and editing, Funding acquisition,
Supervision.

\newpage{}

\bibliographystyle{elsarticle-harv}
\bibliography{reference}

\end{document}